# Hopf Bifurcation in Mean Field Explains Critical Avalanches in Excitation-Inhibition Balanced Neuronal Networks: A Mechanism for Multiscale Variability

Junhao Liang [1,2], Tianshou Zhou [2] and Changsong Zhou[1,3*]

[1]Department of Physics, Centre for Nonlinear Studies and Beijing-Hong Kong-Singapore Joint Centre for Nonlinear and Complex Systems (Hong Kong), Institute of Computational and Theoretical Studies, Hong Kong Baptist University, Kowloon Tong, Hong Kong

[2]School of Mathematics and Guangdong Key Laboratory of Computational Mathematics, Sun Yat-sen University, Guangzhou, P. R. China

[3]Department of Physics, Zhejiang University, 38 Zheda Road, Hangzhou, China

\* Correspondence:  cszhou@hkbu.edu.hk

## Abstract

Cortical neural circuits display highly irregular spiking in individual neurons but variably sized collective firing, oscillations and critical avalanches at the population level, all of which have functional importance for information processing. Theoretically, the balance of excitation and inhibition inputs is thought to account for spiking irregularity and critical avalanches may originate from an underlying phase transition. However, the theoretical reconciliation of these multilevel dynamic aspects in neural circuits remains an open question. Herein, we study excitation-inhibition (E-I) balanced neuronal network with biologically realistic synaptic kinetics. It can maintain irregular spiking dynamics with different levels of synchrony and critical avalanches emerge near the synchronous transition point. We propose a novel semi-analytical mean-field theory to derive the field equations governing the network macroscopic dynamics. It reveals that the E-I balanced state of the network manifesting irregular individual spiking is characterized by a macroscopic stable state, which can be either a fixed point or a periodic motion and the transition is predicted by a Hopf bifurcation in the macroscopic field. Furthermore, by analyzing public data, we find the coexistence of irregular spiking and critical avalanches in the spontaneous spiking activities of mouse cortical slice *in vitro*, indicating the universality of the observed phenomena. Our theory unveils the mechanism that permits complex neural activities in different spatiotemporal scales to coexist and elucidates a possible origin of the criticality of neural systems. It also provides a novel tool for analyzing the macroscopic dynamics of E-I balanced networks and its relationship to the microscopic counterparts, which can be useful for large-scale modeling and computation of cortical dynamics.

# 1. Introduction

The mammal brain consists of tens of billions of neurons, which process information and communicate through electrophysiological action potentials, also known as spikes. This large number of neurons exhibit diverse spiking behaviors across broad ranges of spatial and temporal scales (Okun et al., 2015; Stringer et al., 2019a, 2019b). Understanding the origin and dynamic mechanism of this complexity is crucial for the advancement of neurobiology, the development of therapies for brain diseases and the future design of brain-inspired intelligent systems.

Two striking features can be simultaneously observed at different levels of cortical neuronal systems: 1) irregularity in spiking times, indicated by seemingly random spiking time that resembles Poisson process (Softky and Koch, 1993; Holt et al., 1996) of individual neurons; 2) variability in population firing rates, manifested in widely observed collective neural activities such as population oscillations (Brunel and Wang, 2003; Herrmann et al., 2004) and critical neural avalanches (Beggs and Plenz, 2003; Gireesh and Plenz, 2008; Friedman et al., 2012; Bellay et al., 2015), etc. Biologically, the spiking irregularity has been proposed to originate from the balance between excitation (E) and inhibition (I) inputs so that spiking of neurons is driven by fluctuations (Shu et al., 2003; Okun and Lampl, 2008; Xue et al., 2014), and has been associated with functional advantages in efficient coding and information processing (Denève and Machens, 2016). The emergence of collective cortical activities originates in the fact that neurons interact through recurrent networks (Abeles, 1991), in which dynamic activities can reverberate. As a result, dynamic correlations arise from structural correlations. In particular, even weak pairwise correlation is sufficient to induce strongly correlated collective network activities (Schneidman et al., 2006). Collective neural activities can emerge with different amplitudes and are often organized as critical avalanches with various sizes (Beggs and Plenz, 2003; Gireesh and Plenz, 2008; Friedman et al., 2012; Bellay et al., 2015; Fontenele et al., 2019). These avalanches are cascades of activity bursts in neuronal networks. At criticality, the size and duration of avalanches are approximately distributed according to power-laws, with critical exponents satisfying the scaling relation (Friedman et al., 2012; Fontenele et al., 2019). Avalanches in the critical state can maximize the informational complexity and variability, and are thought to have functional advantages in information processing (Kinouchi and Copelli, 2006; Shew et al., 2009, 2011).

Traditional mean-field theory of E-I balanced networks (Van Vreeswijk and Sompolinsky, 1996, 1998; Renart et al., 2010) with binary neuron and instantaneous synapse explains the spiking irregularity of individual neurons. However, it fails to account for collective neural activities, because it predicts an asynchronous dynamic state with vanishing correlation in unstructured (i.e. random topology) networks. Such vanishing correlation arises due to sparse network connectivity (Van Vreeswijk and Sompolinsky, 1998) or shared excitatory and inhibitory inputs cancelling correlation in dense recurrent networks (Renart et al., 2010). In terms of rate coding, the asynchronous state is not efficient for information processing, as the population firing rate only exhibits a linear response to the input rate (Van Vreeswijk and Sompolinsky, 1996) but without firing rate variability on faster time scales. In this case, the whole network acts as a rate unit for computation. The traditional E-I balanced theory can be generalized in two directions, which are biologically more plausible: structured networks and synaptic kinetics. Firstly, heterogeneous neural network structures (Landau et al., 2016) can induce firing rate variability. For example, clustered network structures (Litwin-Kumar and Doiron, 2012) can induce slow firing rate oscillations and show stimulus-induced variability reductions. Hierarchical modular networks (Wang et al., 2011) can support self-sustained firing rate oscillations across different levels. Spatial networks involving a distance-dependent coupling rule can unveil the distance-tuned correlation relation (Rosenbaum et al., 2017; Darshan et al., 2018) and the emergence of propagating waves (Keane and Gong, 2015; Keane et al., 2018; Gu et al., 2019; Huang et al., 2019) observed in experiments. Secondly, even in unstructured networks, network firing rate oscillations can be induced by realistic synaptic filtering kinetics (Brunel and Wang, 2003; Yang et al., 2017). Such oscillations typically occur in cases where the synaptic decay time scales of inhibition are slower than excitation, which is actually a biologically plausible situation if the synaptic receptors under consideration are AMPA for excitation and GABA for inhibition (Salin and Prince, 1996; Zhou and Hablitz, 1998). More importantly, network oscillation can be sparsely participated by subgroups of neurons, thus preserving the irregular spiking feature of individual neurons (Brunel, 2000; Brunel and Hakim, 2008). Note, however, that theoretical analysis of the macroscopic dynamics of E-I neural network with synaptic kinetics is very difficult. Existing theory is very limited, e.g., by requiring very special assumptions, such as the Lorenz distribution of certain parameters (Dumont and Gutkin, 2019).

Critical avalanches can also rise from neural dynamics under unstructured network topology (Beggs and Plenz, 2003; Kinouchi and Copelli, 2006), while its dynamic origin is still controversial. Previous theories have suggested that critical neural avalanches may arise at the edge of a phase transition. Early studies indicated that it may occur between a quiescent and active phase from critical branching processes (Beggs and Plenz, 2003; Haldeman and Beggs, 2005), while later experimental (Fontenele et al., 2019) and theoretical (di Santo et al., 2018; Dalla Porta and Copelli, 2019) studies also proposed that it may occur near the onset of synchrony. Mechanisms other than criticality that generate avalanches with power-law distribution have also been proposed (Martinello et al., 2017; Touboul and Destexhe, 2017; Wilting and Priesemann, 2019b). The emergence of critical avalanches has also been proposed to be closely related to the maintenance of E-I balance (Lombardi et al., 2012; Poil et al., 2012; Yang et al., 2012). Nevertheless, the exact relationship between neural criticality and E-I balance remains poorly understood.

Here, we try to address the above important open questions. In particular, how E-I balance induced irregular spiking reconciles with collective neural activities? E-I neural networks can organize into a 'sparse synchrony' state (Brunel, 2000; Brunel and Hakim, 2008) where neurons are remained fluctuation-driven to spike irregularly whereas firing rate oscillation emerges in population level. However, the relationship between E-I balance and sparse synchrony is less clear. Most importantly, the mechanism by which E-I balance induced irregular spiking can coexist with critical neural activities (Bellay et al., 2015) in recurrent neural circuit remains unclear. In this work, we first re-examine the dynamics of integrate-and-fire (IF) E-I neuronal network with realistic synaptic kinetics, which can manifest individually irregular spiking activities with different synchronous levels. The network firing rate dynamics can be effectively captured by a set of macroscopic field equations derived by a novel semi-analytical mean-field theory. An advantage of our theory is that it is simple and does not require special properties of the model. The synchronous transition point where network firing rate oscillation emerges is predicted by a Hopf bifurcation in the field equations. We find that critical microscopic avalanche dynamics emerges near the onset of synchronization, with critical exponents approximately satisfying the scaling relations, which manifests the hallmark of criticality (Sethna et al., 2001). The mechanism of critical avalanches could be understood as demographic noise-driven random walks near a macroscopic bifurcation point. On this basis, we propose that the E-I balanced state in the microscopic spiking network corresponds to a stable macroscopic state in the field equations. The asynchronous state, consistent with the traditional theory, corresponds to a stable fixed point, which can be destabilized through a Hopf bifurcation, giving rise to a stable limit circle, corresponding to network firing rate oscillation. As such, the E-I balanced state can incorporate network oscillation with different synchronous levels, which accounts for the coexistence of variability in both individual and population scales. Finally, we empirically verify the coexistence of irregular spiking and collective critical avalanches in the public experimental data of spontaneous spiking activities recorded in mouse somatosensory cortex *in vitro* (Ito et al., 2016). Scaling relations similar to the network model are also found to hold in these critical data sets. Our own analysis further indicates the universality of the observed phenomena in model networks. The theory proposed here explains how collective neural activities coexist with irregular neuron spiking and reveals a possible origin of criticality in neural systems. It also serves as a novel tool to study the dynamics of IF networks with biologically realistic synaptic filtering kinetics, and thus has useful application in large-scale modeling of brain networks.

## 2. Materials and Methods

### 2.1 Spiking neuronal network

We study a leaky IF spiking neuronal circuit. Neurons are coupled by a random network with density $p$ and size $= N_E + N_I$, which consists of $N_E$ excitatory (E) neurons and $N_I$ inhibitory (I) neurons. Thus, each neuron in the network has on average $n_E = pN_E$ E neighbors and $n_I = pN_I$ I neighbors. Each neuron also receives $n_o$ excitatory inputs modelled by independent Poisson processes with frequency $Q_o$, mimicking external inputs for the circuit under consideration. We set $p = 0.2$, $N_E:N_I = 4:1$, $n_o = n_E$ and the network size is $N = 10^4$, unless otherwise specified. The sub-threshold membrane potential of neuron $i$ at time $t$, denoted as $V_i(t)$, is governed by

$$\frac{dV_i}{dt} = f_\alpha(V_i) + J_{\alpha o} \sum_{j \in \partial_i^o} F^E * s_j(t - \tau_l^E) + J_{\alpha E} \sum_{j \in \partial_i^E} F^E * s_j(t - \tau_l^E) \qquad (1)$$

$$+ J_{\alpha I} \sum_{j \in \partial_i^I} F^E * s_j(t - \tau_l^I) \ ,$$

where, $V_i(t)$ is the membrane potential of neuron $i$ (belonging to type $\alpha = E, I$) at time $t$. $\partial_i^\alpha$ represents the $\alpha$ neighbors of neuron $i$. The input sources for neurons include excitatory inputs from external neurons (population $O$), inputs from recurrent excitatory neurons (population $E$) and inputs from recurrent inhibitory neurons (population $I$). The first term of Equation (1) describes the leaky current $f_\alpha(V_i) = (V_{rest}^\alpha - V_i)/\tau_\alpha$, which has the effect to drive the membrane potential back to the leaky potentials, which are set to be $V_{rest}^E = V_{rest}^I = -70\ mV$. The membrane time constants are set as $\tau_E = 20\ ms$, $\tau_I = 10\ ms$ for E and I neurons, respectively. The second to fourth terms of Equation (1) are the external, excitatory recurrent and inhibitory recurrent currents, respectively. Input currents are the summations of the filtered pulse trains. Here, $s_j(t) = \sum_n \delta(t - t_j^n)$ denotes the spike train of the $j$-th neuron. The excitatory and inhibitory synapses have latency period (delay) $\tau_l^E$ and $\tau_l^I$ respectively. For the numerical results presented, we consider $\tau_l^E = \tau_l^I = 0$ for simplicity (i.e. no transmission delay), which is a reasonable approximation for local circuits. The synaptic filter is modelled as a bi-exponential function, i.e. Equation (2).

$$F^\alpha(t) = \frac{1}{\tau_d^\alpha - \tau_r}\left[\exp\left(-\frac{t}{\tau_d^\alpha}\right) - \exp\left(-\frac{t}{\tau_r}\right)\right], t \geq 0.  \quad (2)$$

In Equation (2), we set the synaptic rise time $\tau_r = 0.5\ ms$ for both E and I neurons, while the synaptic decay times $\tau_d^E, \tau_d^I$ depend on the type of presynaptic neuron. We set $\tau_d^E = 2\ ms$ and let $\tau_d^I$ change from $1 \sim 4.5\ ms$ to study the effect of different E and I synaptic filtering time scales. Hence, in our study here, $\tau_d^I$ serves as a control parameter to induce the dynamical transition. Biologically, the inhibition decay time $\tau_d^I$ depends on the constitution of synaptic receptors (Salin and Prince, 1996; Zhou and Hablitz, 1998), and can also be changed by chemicals such as narcotics (Brown et al., 2010). We point out that suitable changes of other model parameters such as synaptic strength, network connection density, etc. may also induce similar dynamical transitions we are going to study below. The integration dynamics is as follows. When the membrane potential reaches the threshold $V_{th} = -50\ mV$, a spike is emitted and the membrane potential is reset to $V_{reset} = -60\ mV$. Then, synaptic integration is halted for 2 ms for E neurons and 1 ms for I neurons, modelling the refractory periods in real neurons. Synaptic weights are set as $J_{EO} = 0.45\ mV$, $J_{IO} = 0.72\ mV$, $J_{EE} = 0.36\ mV$, $J_{IE} = 0.72\ mV$, $J_{EI} = -0.81\ mV$ and $J_{II} = -1.44\ mV$, which will satisfy the balanced condition. Network dynamics are simulated by a modified second-order Runge-Kutta scheme (Shelley and Tao, 2001) with a time step of $dt = 0.05\ ms$. For each parameter, the network is simulated for 16 s with the first 1 s discarded to avoid a transient effect. The statistical indexes are then computed by averaging the results of 15 trials with randomly distributed initial membrane potentials. The dynamics we considered here are current based. The case of conductance-based dynamics where the input of a neuron depends on its membrane potential is further studied in Supplementary Material: Appendix 1. Model Extensions.

## 2.2 Mean-field theory of network dynamics

### 2.2.1 Derivation of the macroscopic field equations.

We will develop a semi-analytical mean-field theory to approximate the average dynamics of the network by noting the fact that 1) the network topology is homogeneous and 2) the number of neuron $N$ is large. The following mean-field derivation holds for large enough $N$ and the network size is explicitly included in the field model equations (see Equations (11), (12) below).

We denote the average membrane potential of the network as $V_\alpha = \langle V_i \rangle_{i \in \alpha} := \langle V_i \rangle_\alpha$, $\alpha = E, I$. The goal is to derive a set of self-consistent field equations governing the temporal dynamics of $V_\alpha$.

First, by noting that the convolution $F^\alpha * s_j(t - \tau_l^\alpha)$ obeys the equation

$$\left(\tau_d^\alpha \frac{d}{dt} + 1\right)\left(\tau_r \frac{d}{dt} + 1\right)[F^\alpha * s_j(t - \tau_l^\alpha)] = \sum_n \delta(t - t_j^n - \tau_l^\alpha), \quad (3)$$

then we have

$$\left(\tau_d^\alpha \frac{d}{dt} + 1\right)\left(\tau_r \frac{d}{dt} + 1\right)\langle \sum_{j \in \partial_i^\alpha} F^\alpha * s_j(t - \tau_l^\alpha) \rangle_{i \in E, I} = \langle \sum_{j \in \partial_i^\alpha} \sum_n \delta(t - t_j^n - \tau_l^\alpha) \rangle_{i \in E, I}. \quad (4)$$

Under mean-field approximation, each neuron is the same in terms of their neighbors, so that

$$\langle \sum_{j \in \partial_i^\alpha} \sum_n \delta(t - t_j^n - \tau_l^\alpha) \rangle_{i \in E, I} \approx n_\alpha Q_\alpha(t - \tau_l^\alpha), \quad (5)$$

where $n_\alpha$ is the average number of $\alpha$ neighbors of a neuron in the network and $Q_\alpha(t)$ is the mean firing rate of $\alpha$ type neurons, defined as

$$Q_\alpha(t) = \lim_{\Delta t \to 0} \frac{1}{\Delta t} \int_t^{t+\Delta t} \langle \sum_n \delta(s - t_j^n) \rangle_{j \in \alpha} ds. \quad (6)$$

In the standard definition of firing rate (Dayan and Abbott, 2001) of a neuron, the average in Equation (6) is taken over different simulation trials. Since ergodicity and the network homogeneity, neurons within a population should have the same firing rate and it can be computed through Equation (6) (i.e. population averages can be thought as sample averages) when the network is large enough in the stationary state. Formally, for measuring the firing rate from data, the time interval $\Delta t$ in Equation (6) has to be finite. Here we choose $\Delta t = 1\ ms$ (shorter than the refractory period) so that a neuron can at most have one spike between $t$ and $t + \Delta t$. Then, by the definition of $\delta$ function, $Q_\alpha(t)$ represents the proportion of $\alpha$ type neurons that spike between $t$ and $t + \Delta t$ as well as the mean firing rate of $\alpha$ type neurons at time $t$ with unit per *ms*.

In previous analysis framework of IF neurons through continuous stochastic process theory (Burkitt, 2006), the membrane potential $V_i$ of neuron $i$ cannot cross the spiking threshold ($V_i$ is restricted to $(-\infty, V_{th})$ with $V_{th}$ being an absorbing boundary). This is a theoretical artefact, contrary to the true neurophysiology. Furthermore, in numerical integration, the resetting is achieved by finding those neurons whose membrane potential increases over the spiking threshold in each numerical step (Shelley and Tao, 2001). This inspires us to naturally consider that a neuron $j$ should have a spike at time $t$ if $V_j(t) > V_{th}$. Formally, we can consider $V_j(t)$ as the membrane potential of neuron $j$ at time $t$ before the resetting rule in each numerical step, then

$$Q_\alpha(t) = \langle H(V_j - V_{th}) \rangle_{j \in \alpha}, \quad (7)$$

where $H$ is the Heaviside function $H(x) = \begin{cases} 1, x \geq 0 \\ 0, x < 0 \end{cases}$. Equation (7) explicitly builds the link that the population firing rate is the proportion of the neurons whose membrane potential is above the spiking threshold. As a preliminary approximation, we assume $V_j(t)$ obey a Gaussian distribution $P_\alpha(V)$ with time-dependent mean $V_\alpha(t)$ and time-independent variance $\sigma_\alpha^2$. We will verify that this assumption is plausible in the network and dynamic regimes we studied, referring to **Figure 1D** below. Then,

$$Q_\alpha(t) = \langle H(V_j - V_{th})\rangle_{j\in\alpha} = \int_{V_{th}}^{\infty} P_\alpha(V)dV = \frac{1}{2} - \frac{1}{2}\text{erf}(\frac{V_{th}-V_\alpha}{\sqrt{2}\sigma_\alpha}), \tag{8}$$

where $erf$ is the error function $erf(x) = \frac{2}{\sqrt{\pi}}\int_0^x e^{-t^2}dt$. Although there is no elementary expression for the error function, it can be approximated by elementary functions. For example, a good approximation that can keep the first and second moments is $erf(x) \approx \tanh(\frac{\pi x}{\sqrt{6}})$. Under this approximation, we have

$$Q_\alpha(t) = \frac{1}{2} - \frac{1}{2}\tanh\left(\frac{\pi}{\sqrt{6}}\frac{V_{th}-V_\alpha}{\sqrt{2}\sigma_\alpha}\right) = \frac{1}{1+\exp(\frac{V_{th}-V_\alpha}{\sigma_\alpha}\frac{\pi}{\sqrt{3}})}. \tag{9}$$

Here, the standard deviation of the voltage, $\sigma_\alpha$, acts as an effective parameter to construct the voltage-dependent temporal firing rate. Note that this approximation scheme basing only on the first-order statistics neglects several factors that affect the accurate firing rate, including higher order statistics, noise correlation and refractory time. Thus, it does not have an analytical form and should be estimated numerically, from the steady-state mean voltage $V_\alpha^{ss}$ and mean firing rate $Q_\alpha^{ss}$ and Equation (9) that

$$\sigma_\alpha = \frac{V_{th}-V_\alpha^{ss}}{\ln[(Q_\alpha^{ss})^{-1}-1]}\frac{\pi}{\sqrt{3}}. \tag{10}$$

The sigmoid transfer function Equation (9) is the intrinsic nonlinear property that induces oscillation transition in the field model. Note that neural field models of Wilson-Cowan type (Wilson and Cowan, 1972) would also contain a presumed sigmoid transfer function. Field models of this type can also qualitatively reproduce some dynamic features of E-I neuronal networks. Here we explicitly construct the sigmoid function from the microscopic spiking network. Thus, the quality of the scheme depends on suitable choices of effective parameters $\sigma_\alpha$ and once $\sigma_\alpha$ are chosen as suitable values, our field equations can predict the dynamics of the E-I network quantitatively (see Supplementary Material: Appendix 2. Sensitivity of the Critical Points on the Effective Parameters, where we study the sensitivity of the oscillation transition with respect to the effective parameters $\sigma_\alpha$).

Denote $\Phi_\alpha(t) = \langle \sum_{j\in\partial_i^\alpha} F^\alpha * s_j(t - \tau_l^\alpha)\rangle_{E,I}$ as the averaged synaptic time course of $\alpha$ inputs received by a neuron and from Equations (4), (5), (9) it will obey

$$\left(\tau_d^\alpha \frac{d}{dt} + 1\right)\left(\tau_r \frac{d}{dt} + 1\right)\Phi_\alpha = \frac{n_\alpha}{\left[1+\exp\left(\frac{V_{th}-V_\alpha(t-\tau_l^\alpha)}{\sigma_\alpha}\frac{\pi}{\sqrt{3}}\right)\right]}, \quad \alpha = E, I. \tag{11}$$

Note that each neuron receives $n_o$ independent Poisson spike trains externally with rate $Q_o$. Thus, the input of each neuron has a variance $n_o Q_o$. If we do not consider its filtering effect since the fast decay time of excitatory synapse, by diffusion approximation we know the external input $\sum_{j \in \partial_i^o} F^E * s_j(t - \tau_l^E)$ of each neuron can be approximated by $n_o Q_o + \sqrt{n_o Q_o} \xi_i(t)$, where $\xi_i(t)$ is a standard Gaussian white noise (GWN) with zero mean and unit variance. Since $\{\xi_i(t)\}_i$ are independent GWNs, $\langle \sqrt{n_o Q_o} \xi_i(t) \rangle_\alpha$ can be equivalently approximated by $\sqrt{n_o Q_o / N_\alpha} \xi_\alpha(t)$, where $\xi_\alpha(t)$ is another standard GWN. Thus, taking the average $\langle . \rangle_\alpha$ of the original Equation (1) and note that in the leaky IF model, $f_\alpha$ is linear, we arrive at

$$\frac{dV_\alpha}{dt} = f_\alpha(V_\alpha) + J_{\alpha o}\left(n_o Q_o + \sqrt{\frac{n_o Q_o}{N_\alpha}} \xi_\alpha(t)\right) + J_{\alpha E}\Phi_E + J_{\alpha I}\Phi_I, \alpha = E, I . \quad (12)$$

In the field Equation (12), $\xi_E(t)$ and $\xi_I(t)$ are two independent GWNs. We find that this approximation is independent of the nature of noise in the spiking network model. In the network model Equation (1), the nature of noise from external inputs is synaptic-filtered Poisson shot noise. We further examine the case where external inputs in Equation (1) are with GWNs and the case of constant external input (i.e. no noise) and find that in such cases Equation (12) can still well approximate the macroscopic dynamics of the network (for constant external input the network dynamics is not stochastic, but the spiking activity still appears irregular due to the chaotic nature of the network (Van Vreeswijk and Sompolinsky, 1996, 1998)). Generally speaking, the mean-field theory only holds when the system size is infinity. The incorporation of noise into the field model can smooth out the systematic errors, compensate the finite size effect and make it closer to the true rate dynamics statistically. Thus, for numerical simulation of the field equations we will keep the noise terms in Equation (12) whereas the deterministic counterpart would be used for stability analysis.

In summary, we have proposed a novel technique to derive a set of self-consistent field Equations (11), (12), to approximate the average dynamics of the original spiking network Equations (1), (2).

### 2.2.2 Analysis of the steady-state dynamics.

The deterministic steady-state (fixed point) of the field model can be found from Equations (11), (12) by letting $\frac{d}{dt} = 0$ and assuming $\xi_\alpha(t) = 0$, resulting in algebraic equations

$$f_\alpha(V_\alpha) + J_{\alpha o} n_o Q_o + J_{\alpha E} n_E Q_E + J_{\alpha I} n_I Q_I = 0, \alpha = E, I . \quad (13)$$

where $Q_E, Q_I$ are given by Equation (9). In our case, the synaptic strengths and external input rates are the major parameters determining the value of the fixed point, while synaptic decay time would affect its stability. This is because the value of steady-state does not depend on $\tau_d^\alpha, \tau_r$, which can also be seen from Equation (2) that the synaptic filter is normalized ($\int_0^\infty F^\alpha * \delta(t) = 1$ independent of $\tau_d^\alpha, \tau_r$) so that synaptic rise and decay times would not affect the time-averaged firing rate. Thus, the scheme here cannot capture the nontrivial effect of synaptic filtering on affecting the firing rate (e.g. see the formula given by Fourcaud and Brunel (Fourcaud and Brunel, 2002)). Note that in general settings of balanced networks with dense and strong coupling (Renart et al., 2010), the quantities $n_o, n_E, n_I$ are of order $O(N)$ but the synaptic weights are of order $O(N^{-1/2})$. When $N$ is large enough, the first term in Equation (13) can be neglected and it reduces to

$$J_{\alpha o} n_o Q_o + J_{\alpha E} n_E Q_E + J_{\alpha I} n_I Q_I = 0, \alpha = E, I, \quad (14)$$

which is a set of linear equations to solve the steady firing rate $Q_E, Q_I$. To guarantee a unique positive solution in this case, the sequence $\{\frac{J_{EO}}{J_{IO}}, \frac{J_{EI}}{J_{II}}, \frac{J_{EE}}{J_{IE}}\}$ should be in ascending or descending order, which is the so-called balanced condition in the traditional theory (Van Vreeswijk and Sompolinsky, 1996, 1998; Renart et al., 2010). Thus, the theory here is a generalization of the traditional theory of balanced network.

In the traditional theory of asynchronous dynamics, the E-I balanced state can be considered as the existence of a stable fixed point of Equations (11), (12). Now we can consider how the stability of this fixed point can be changed with the aid of this dynamic form. For the case without synaptic delay (i.e. $\tau_l^E = \tau_l^I = 0$), the field equations are ordinary differential equations. By taking $X = \left(V_E, V_I, \Phi_E, \frac{d\Phi_E}{dt}, \Phi_I, \frac{d\Phi_I}{dt}\right)^T$, the field model Equations (11), (12) can be written in the first-order form $\frac{dX}{dt} = F(X)$ without considering the noise. The Jacobian matrix at the steady state is

$$J = \begin{pmatrix} -\frac{1}{\tau_E} & 0 & J_{EE} & 0 & J_{EI} & 0 \\ 0 & -\frac{1}{\tau_I} & J_{IE} & 0 & J_{II} & 0 \\ 0 & 0 & 0 & 1 & 0 & 0 \\ \frac{n_E Q_E'(V_E)}{\tau_d^E \tau_r} & 0 & -\frac{1}{\tau_d^E \tau_r} & -\frac{1}{\tau_d^E} - \frac{1}{\tau_r} & 0 & 0 \\ 0 & 0 & 0 & 0 & 0 & 1 \\ 0 & \frac{n_I Q_I'(V_I)}{\tau_d^I \tau_r} & 0 & 0 & -\frac{1}{\tau_d^I \tau_r} & -\frac{1}{\tau_d^I} - \frac{1}{\tau_r} \end{pmatrix}, \quad (15)$$

with $Q_\alpha'(V_\alpha) = \frac{\pi \exp[(V_{th} - V_\alpha)\pi/(\sqrt{3}\sigma_\alpha)]}{\sqrt{3}\sigma_\alpha \left(1 + \exp\left[\frac{(V_{th} - V_\alpha)\pi}{\sqrt{3}\sigma_\alpha}\right]\right)^2}$ estimated at the steady-state value of $V_\alpha$ given by Equation (13). The eigenvalues of $J$ can determine the stability of the steady state. Note that many models use single exponential function as the synaptic filter, i.e. $\tau_r = 0$, and in this case the dynamic form becomes 4-dimensional with $X = (V_E, V_I, \Phi_E, \Phi_I)^T$. For models without considering synaptic filtering effect (that is, the case of instantaneous synapse where $\tau_r = \tau_d^\alpha = 0$), the dynamic form becomes 2-dimensional, which can be considered as the dynamic form of traditional theory without synaptic kinetics. In these cases, the stability analysis can be performed in a similar way. In the presence of synaptic transmission delay, the field equations would become delay differential equations. In this case, stability analysis would in general become more difficult.

## 2.3 Statistical analysis of model and experiment data

### 2.3.1 Spike count series.

For statistically analyzing neural dynamics, the neuron spike train series have to be constructed from the model simulation data or the experimental data as follows. The time axis is first divided into consecutive time windows with sizes $\Delta t$ ms. The number of spikes of neuron $i$ are then counted in each window to obtain a discrete sequence $N_i(t)$, which is designated as the spike count series of neuron $i$ with time windows $\Delta t$. Alternatively, the number of spikes of the whole neuron population can be counted in each window. This constructs the population spike count series $N_\alpha(t)$ for the $E$ and $I$ populations, respectively. Furthermore, $q_\alpha(t) = \frac{N_\alpha(t)}{n_\alpha \Delta t}$ is the population averaged firing rate series and the power spectrum of the population firing rate series can indicate the collective oscillation property. For computing certain quantities, such as the correlation between neurons, the spike count series is filtered by a square kernel $K_T(t) = \begin{cases} \frac{1}{T}, t \in [-T, 0] \\ 0, other \end{cases}$ with length $T$ and the ensuing filtered spike train is defined as $\widetilde{N}_i(t) = K_T * N_i = \frac{1}{T}\sum_{s=0}^{T-1} N_i(t-s)$.

### 2.3.2 Quantifying spiking irregularity and firing rate variability.

The spiking time irregularity of a neuron can be quantified using the CV (coefficient of variance), which is defined as the standard deviation of the neuron ISIs (inter-spike intervals) over its mean. Totally regular activities have CV values of 0, while Poisson processes have CV values of 1. A higher CV value indicates larger irregularity. We measure the firing rate variability on the individual neuron level and the population level by the relevant Fano Factor (FF). The FF of neuron $i$ is defined as $FF_i = var(N_i(t))/\langle N_i(t)\rangle$. The average and the variance here are taken across the spike count series. Similarly, population FF is defined as $FF_\alpha = var(N_\alpha(t))/\langle N_\alpha(t)\rangle$. Note that FF depends on the time window size $\Delta t$ to construct the spike count series (we will use $\Delta t = 50\ ms$). The population FF can also be computed using the macroscopic field equations, as the field equations predicts the population firing rate, which can be transferred to population spike counts (by multiplying the number of neurons in the population).

### 2.3.3 Quantifying network synchrony.

The synchrony of the network can be characterized from two aspects: the cross-correlation of spiking times and the coherence of the membrane potential (Golomb, 2007). The former quantifies the coherence of threshold events whereas the later quantifies the coherence of the subthreshold dynamics.

We employ the commonly used Pearson correlation coefficient (PCC) to quantify the synchrony of the spiking time. The spike count series of neuron $i$ is first constructed with time window $\Delta t =$ $1\ ms$ and then filtered by a square kernel with length $T = 50\ ms$. The PCC between neuron $i$ and $j$ is defined as $c_{ij} = \frac{cov(n_i(t), n_j(t))}{\sqrt{var(n_i(t))var(n_j(t))}}$. The details such as filtering in calculating PCC would affect its exact value (Cohen and Kohn, 2011), but not the qualitative change. The index $\langle c_{ij} \rangle_{i,j \in E}$, PCC averaged over all excitatory neuron pairs, is used to quantify the network synchrony degree of threshold events.

The voltage series of each neuron is constructed for each millisecond that $V_i(k) := V_i(t)|_{t=k\ ms}$. Voltage coherence is defined as $= \sqrt{\sigma_\alpha^2 / \langle \sigma_i^2 \rangle_\alpha}$, where $\sigma_\alpha^2 = \langle V_\alpha^2 \rangle_t - \langle V_\alpha \rangle_t^2$ is the variance of the mean voltage $V_\alpha = \langle V_i \rangle_{i \in \alpha}$ and $\sigma_i^2 = \langle V_i^2 \rangle_t - \langle V_i \rangle_t^2$ is the variance of the voltage of neuron $i$. The voltage coherence of the excitatory population is used to quantify the coherence of the subthreshold dynamics.

We further use the CV of the excitatory population firing rate series constructed with $\Delta t = 1\ ms$ to quantify temporal firing rate variability at short timescales, which is another way of indicating the network synchrony, as stronger synchrony indicates larger population firing rate variability at short timescales.

**2.3.4 Neuronal avalanche analysis.**

We measure the neuronal avalanches of the excitatory neuron population from its population spike count series $N_E(t)$ constructed with window (bin) size $\Delta t$. An avalanche is defined as a sequence of consecutive non-empty bins, separated by empty bins (with no spiking inside). The size $S$ of an avalanche is defined as the total number of spikes within the period and the duration $T$ is defined as the number of time bins it contains. To compare different data sets in a unified standard, window size is chosen as the average ISI of the merged spiking train (constructed by merging the spike trains of all neurons). Thus, it depends on the mean firing rate of neurons in different data sets. This choice has been described as the 'optimal' window size to measure avalanches (Beggs and Plenz, 2003), as excessively small or large windows would lead to systematic bias. A further advantage of this choice is that the measured size and duration would approximately lie in the same scale ranges, allowing better comparison. Note that the definition of avalanche here is the time-binning (non-causal) avalanche, which corresponds to the case of experimental measurement (Beggs and Plenz, 2003; Bellay et al., 2015), but is different from the causal avalanche (Williams-Garcia et al., 2017) studied by many physical models rooted in critical branching processes.

Avalanche size obeying a power-law distribution is a typical sign of scale-invariance, which is a statistical property. We use a simple method to quantify how far the avalanche size distribution deviates from a power-law distribution. The avalanche size frequency distribution histogram $(S, P(S))$, is first obtained with 80 plotting-bins from minimum to maximum size. The least squares method is then used to find the best-fit-line in doubly logarithmic coordinates, such that $\sum_S [lgP(S) - (b_0 + b_1 lgS)]^2$ is minimised. After obtaining the best-fit coefficients $(b_0, b_1)$, the fitted frequency distribution values were estimated as $P_{fit}(S) = 10^{b_0 + b_1 lgS}$. Finally, the normalised distance, defined as $D = \sum_S S|P(S) - P_{fit}(S)| / \sum_S SP(S)$, the average size difference per avalanche between the actual and fitting frequency distribution normalized by the mean avalanche size, is used to measure the distance to the best fitting power-law distribution. This distance acts as an index to indicate the degree of criticality for comparison.

The maximum likelihood estimation (MLE) method using the NCC toolbox (Marshall et al., 2016) is used to estimate the critical exponents. The toolbox provides a doubly truncated algorithm based on the MLE to find the range that passed the truncation based KS statistics test (Marshall et al., 2016). This truncation scheme can avoid the noise (in the small avalanche size range) and finite size effect (in the large avalanche size range) interruption in estimating the critical exponents. We find the largest truncated range that can pass the KS-based test with a *p* value larger than 0.1 and boarder than one-third of the whole range on the logarithm scale. This means that the data can produce a KS-statistics value that is less than the values generated by at least 10% of the power-law models in the truncated range. The estimated slopes within the truncated ranges in the avalanche size and duration distributions define the critical exponents $P(S) \sim S^{-\tau}$ and $P(T) \sim T^{-\alpha}$. A third exponent is defined as $\langle S \rangle(T) \sim T^{1/\sigma vz}$, where $\langle S \rangle(T)$ is the average size of avalanches with the same duration $T$. This exponent is fitted using a weighted least squares method (Marshall et al., 2016) to those avalanches that fall into the truncated duration range for estimating $\alpha$.

### 2.3.5 Experimental data analysis.

We used the public experimental data measuring the neuronal spiking activity in mouse somatosensory cortex cultures *in vitro* (Ito et al., 2016). A total of 25 data sets were used and the length of each record is one hour, with the exception of Set 19, which was 48 minutes long. The recordings were performed on organo-typical slice cultures after two to four weeks *in vitro*, without stimulation (Ito et al., 2014). Spiking times were sorted with a PCA-based algorithm (Litke et al., 2004) to locate the signals recorded with a large and dense multi-electrode array of 512 electrodes.

Under *in vitro* conditions, spiking in the culture clearly shows up-down state transition. Active spiking periods (up-state) and silent periods (down-state) alternated slowly with a frequency of *circa* 0.1 Hz. We focus on analyzing the up-state defined as follows. For each data set, the population firing rate series was first constructed with a time window of $\Delta t = 10\ ms$ and then filtered by a square kernel with length $T = 100\ ms$. Then, the up-state is defined as the time periods that the population firing rates are higher than 30% of the maximum rate of the given dataset and last longer than 1s. We further examine the power spectral density of the population firing rate series (before further filtering) in the up-state, the distribution of firing rates of the neurons, CV of ISIs and the avalanches in the up-states. The time bin for measuring the avalanche was the mean ISI, averaged through the up-state, of the merged spike train. To test critical properties, we first determine whether the size and duration distribution is close to a power-law and then estimate its critical exponents. As in our analysis of modelling data, the doubly truncated and statistical test algorithm from the NCC toolbox (Marshall et al., 2016) is used. We accept the power-law distribution of a data set if the following two conditions can be jointly satisfied: 1) the truncated range has to be boarder than one-third of the whole range in the logarithm scale and; 2) the data in the truncated range can pass the KS-based test with p value larger than 0.1. For the cases deemed to be power-law distribution, the truncated ranges, estimated exponents and $p$ values in the KS-based test are shown. For the data sets that have power-law avalanche distributions in both size and duration, we further compare the critical exponents $1/\sigma\upsilon z$ and the value $\frac{\alpha-1}{\tau-1}$ to see whether scaling relation (Equation (16) below) holds. To test whether power-law distribution is the intrinsic structure of the data, we also analyze the surrogated data constructed by randomly shuffling the ISIs of neurons in each up-state period. This random shuffling can destroy the intrinsic temporal correlation structures of the avalanches.

## 3. Results

### 3.1 Network synchrony arises from loose temporal balance

To begin, we first intuitively illustrate the non-trivial effect of synaptic filtering (**Figure 1A**) in shaping the network dynamics. For fast inhibition decay time, the network spiking dynamics is asynchronous (Renart et al., 2010), as shown in **Figure 1B**. Such a 'strict balance' is the case of traditional E-I balance theory (Van Vreeswijk and Sompolinsky, 1996, 1998) with inhibition domination, where network inhibitory feedback can cancel the excitatory current spill on a fast time scale (**Figure 1H**).

This scenario of strict balance breaks down when inhibition becomes slow, which induces an effective delay in the inhibitory cancellation (**Figure 1I**). Such a delay before the cancellation would induce a window during which excitatory current spills over the network, resulting in the collective spiking shown in **Figure 1C**. The strength of the collective activity depends on the length of this excitation-dominant window. As inhibition becomes dominant again after a delay, a temporal quiescent episode ensues after the collective spiking. Thus, slow synaptic inhibition induces network oscillation, as was also shown in previous studies (Brunel and Wang, 2003; Yang et al., 2017; Huang et al., 2019). In this 'loose balance' scenario, the network maintains balance on a slower time scale. In this case, as excitation and inhibition dominate alternately, the strict balance is temporally broken on a fast time scale. Note that in the loose balance state, the delay in the inhibitory cancellation is merely a few milliseconds (**Figure 1I**), consistent with the inhibition tracking delays observed in experiments (Okun and Lampl, 2008).

Alternatively, these different collective behaviors can be seen from the network population firing rate dynamics. The network firing rate in the strict balance, asynchronous state is almost steady (**Figure 1E**), whereas the loose balance induces firing rate oscillation (**Figure 1F**), with fast firing rate variability at the population level. The emergence of oscillation implies a regularity of the population dynamics of the network. However, it is important to stress that the spiking of the neurons is still highly irregular. There are at least two reasons for this irregularity. First, the population oscillation is quasi-periodic rather than periodic due to the stochastic nature of the network dynamics. Second, and more importantly, each collective burst is randomly participated in by partial neurons, manifested by a firing rate much smaller than the oscillation frequency, referring to **Figure 2E**. This can also be understood from the fact that the firing rate is still low at the peaks of spiking activity corresponding to spiking synchrony (**Figure 1F**). As such, the dynamical phenomenon here with loose balance was also referred to as 'sparse synchrony' (Brunel and Hakim, 2008). The histogram counts of the coefficient of variance (CV) of the inter-spike intervals (ISIs) of neurons (see Materials and Methods) in the network are shown in **Figure 1G**. In both asynchronous and synchronous states, the overall value of the CV is around 1, indicating that the spiking irregularity approximately resembles the Poisson process (Holt et al., 1996). As our mean-field theory assumes that the membrane potentials of the neurons in the network approximate Gaussian, we further analyze the distribution of the membrane potentials. As shown in **Figure 1D**, in both fast and slow inhibition cases the distribution is not totally Gaussian and instead right skewed, a feature of the finite threshold IF dynamic (Brunel, 2000; Keane and Gong, 2015). However, the skewness (defined as the third central moment divided by the cubic of the standard deviation) and kurtosis (defined as the fourth central moment divided by the quartic of the standard deviation) measuring the deviation from Gaussian are (-0.67, 3.12) and (-0.66, 3.3) in the fast and slow inhibition cases respectively, which are not far from (0, 3) for the case of Gaussian distribution. Thus, our theory that assumes a Gaussian distribution is still effective, as will be shown later.

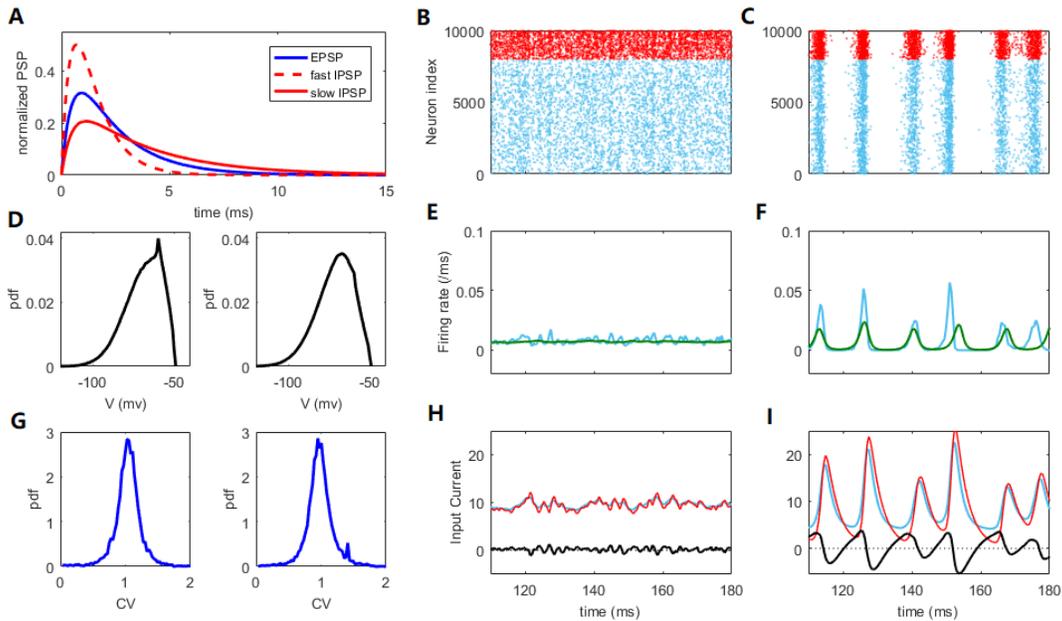

**Figure 1. Synchronization and network oscillation induced by slow inhibition in balanced networks**. Fast inhibition induces strict balance with asynchronous spiking and an almost steady network firing rate. Slow inhibition results in loose balance, with synchronous grouped spiking and network oscillation. Individual neurons spike irregularly in both cases. **(A)** Normalized E/I post-synaptic response when receiving a pre-synaptic spike. **(B, C)** Raster plot of the spiking time. Each blue/red point corresponds to a spike of the E/I neuron. **(D)** The distribution of membrane potential. **(E, F)** The corresponding firing rate of E population (blue) and the firing rate predicted by the field equations (green). **(H, I)** The blue/red curves represent the average input E/I current of a neuron in the network and the black curves represent the difference between them. **(G)** The pdf. of the CV of ISI of E neurons. The parameters are set as $Q_o = 5\ Hz$; $\tau_d^I = 1\ ms$ in the fast inhibition case (**B, E, H** and left panels of **D, G**) and $\tau_d^I = 3.5\ ms$ in the slow inhibition case (**C, F, I** and right panels of **D, G**).

### 3.2 Mean-field theory predicts the synchronous transition

The network dynamic transition induced by a looser E-I balance can be characterized by the dynamics of population firing rate (**Figure 1E, F**). However, it is theoretically challenging to analyze the population dynamics of IF networks with synaptic kinetics. Here, we propose a novel mean-field approximation theory to derive the macroscopic dynamic equations of an IF network, i.e. Equations (11), (12). This technique for deriving the macroscopic field equations is highly generalizable. Extensions to the cases of time-varying inputs and conductance-based dynamics are presented in Supplementary Material: Appendix 1. Model Extensions.

For numerically estimated $\sigma_\alpha$ (i.e. use Equation (10) to compute $\sigma_\alpha$ under different external input strength $Q_o$), the mean firing rate of the network can be correctly estimated from the field Equations (11), (12) (**Figure 2A** blue). We further consider whether the field equations can predict the firing rate with fixed parameters $\sigma_\alpha$. Following the derivation of Equation (12), if we simply assume $\frac{dV_i}{dt} \approx const - \frac{V_i}{\tau_\alpha} + J_{\alpha o}\sqrt{n_o Q_o}\xi_i(t)$, then $V_i$ is Gaussian distributed with standard deviation $\sigma_\alpha = \sqrt{J_{\alpha o}^2 n_o Q_o \tau_\alpha}$. **Figure 2A** green shows the corresponding results of fixed $\sigma_\alpha = \sqrt{J_{\alpha o}^2 n_o Q_o \tau_\alpha}$ for $Q_o = 5\ Hz$. It cannot predict the exact firing rate but can correctly predict the linear response to external input, a property of asynchronous balanced network (Van Vreeswijk and Sompolinsky, 1996).

The dynamic difference between the asynchronous state and synchronous state can already be predicted by the field equations in terms of the population firing rate dynamics, as shown in **Figure 1E, F**. The mechanism is explained by a Hopf bifurcation in the field equations through stability analysis (see Materials and Methods for further details), as shown in **Figure 2B**. In the case of fast inhibition, the fixed point of the field model is generally a stable focus, whose Jacobian has complex eigenvalues with negative real parts. In this case, the network firing rate only fluctuates mildly due to noise perturbations. When $\tau_d^I$ increases, the fixed point will lose its stability through a supercritical Hopf bifurcation, as indicated by a pair of its dominant complex conjugate eigenvalues $\lambda = \alpha + i\omega$ crossing the imaginary axis. The Hopf bifurcation predicts that the stable fixed point will give way to a stable periodic solution, whose amplitude grows from zero. The frequency of the periodic motion can be estimated as $\omega/2\pi$ in the linear order. This prediction is approximately equal to the numerically measured peak frequency of the network excitatory firing rate oscillation near the critical bifurcation point, as shown by the blue circles in **Figure 2B**. Previous studies have shown that E-I networks can undergo a transition to oscillation through perturbation analysis by assuming the form of the network steady-state firing rate solution (Brunel and Wang, 2003; Brunel and Hakim, 2008) and the phenomenon has been associated with a Hopf bifurcation of rate equations with effective transmission delay (Brunel and Hakim, 2008). Here, our mean-field approach with the aid of macroscopic field equations derived from microscopic neuronal network straightforwardly reveals the Hopf bifurcation mechanism during this transition to sparse synchrony state.

As shown in **Figure 2C**, the network synchrony increases dramatically after $\tau_d^I$ crosses the Hopf bifurcation point. **Figure 2D** shows that the network temporal firing rate variability (see Materials and Methods for further details) increases conspicuously during this transition, which is also qualitatively predicted by the field equations. The CV of ISIs averaged over the excitatory population is shown in **Figure 2E**. During the onset of collective oscillation, the averaged CV of ISIs first slightly decreases and then increases (as the strong bursting oscillation activity develops at around $\tau_d^I \approx 4\ ms$, referring to **Figure 3A** bottom panel) while its overall value is around 1. The coexistence of irregular spiking and collective oscillation can be understood from the ratio between the mean firing rate of neurons and peak frequency of network oscillation. If each oscillation is participated by all the neurons, then the ratio should approach one. However, as can be seen from **Figure 2E**, the synchrony is sparse (Brunel and Hakim, 2008) in that this ratio is nearly 0.1 after oscillation onset, implying that each oscillation is randomly participated by only 10% of the neurons. Equivalently, each neuron only spikes sparsely and randomly participates in about 10% of the collective oscillations, giving rise to high variability of the ISI. As $\tau_d^I$ further increases to bursting onset, this ratio rapidly increases to approaching one, implying that almost all the neurons participate in each oscillation in the bursting state, where the inhibitory feedback is too slow such that the excitatory current can spill over the whole network in the excitatory dominant period.

In addition to spiking time variability, we further examine the fluctuations of firing rate in a short time scale (time window $50\ ms$) by Fano Factor (FF) (Teich et al., 1997). As shown in **Figure 2F**, the change of FF of the individual neurons (see Materials and Methods for further details) is similar to that of CV (**Figure 2E**), indicating that the individual firing rate fluctuation does not increase with the onset of collective synchrony. In contrast, the change of population FF (**Figure 2F**, and see Materials and Methods for details) is similar to the spiking correlation (**Figure 2C**) and temporal variability of firing rate (**Figure 2D**). This trend was further qualitatively confirmed by the field model prediction, as shown in **Figure 2F**. Thus, it appeared that the observed collective oscillation is a network property rather than an individual neuron property. Hence, the network collective oscillation activity is compatible with individual neuron irregular spiking. The latter property is the prominent feature of E-I balance (Okun and Lampl, 2009).

In summary, strict balance with asynchronous network spiking is predicted by a stable focus whereas loose balance with sparsely synchronous network spiking is predicted by a stable limit circle in the field equations. Both strict and loose balance conditions support the irregular spiking of individual neurons.

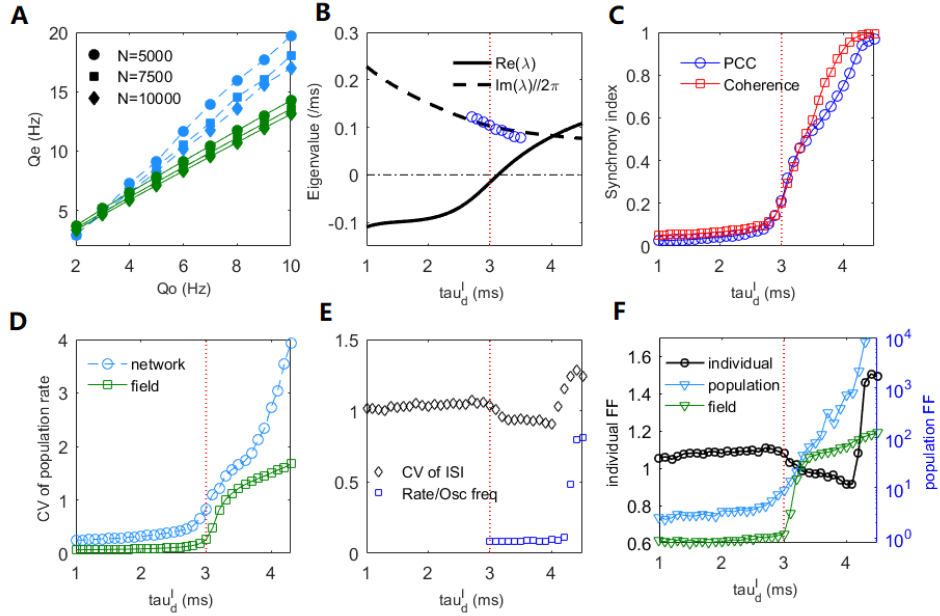

**Figure 2. The mean-field theory prediction of the dynamic transition of the spiking network.** **(A)** The excitatory firing rate in the asynchronous steady-state ($\tau_d^I = 1\ ms$) under different strengths of external input and different network sizes. Blue symbols, obtained by the field equations with numerically estimated $\sigma_\alpha$ under different input rates $Q_o$, fits well with the network simulations (blue dashed lines). Green symbol-lines are obtained by the field equations fixed $\sigma_\alpha = \sqrt{J_{\alpha o}^2 n_o Q_o \tau_\alpha}$ for $Q_o = 5\ Hz$. Synaptic strengths $J_{\alpha\beta}$ are multiplied by 1.41 and 1.15 in the cases of $N = 5000$ and $N = 7500$, respectively to maintain the usual scale $J_{\alpha\beta} \sim N^{-1/2}$. **(B)** Deterministic field equations predict that a Hopf bifurcation occurs with the increase of inhibitory decay time $\tau_d^I$ at a critical value around $\tau_d^I \approx 3\ ms$ (indicated by a vertical dashed line). The real part and imaginary part divided by $2\pi$ of the dominant eigenvalue are denoted by the solid and dashed lines, respectively. Blue circles are peak frequencies of the excitatory firing rate oscillation from network simulation. **(C)** The PCC and coherence index show the emergence of network oscillation as $\tau_d^I$ increases. **(D)** The CV of population firing rate increases with the increase of $\tau_d^I$. **(E)** The average CV of ISIs and the ratio between mean firing rate and peak frequency of network oscillation. **(F)** The average FF of individual neurons (left axis), and the FF of the network population spiking counts in network and field model (right axis). Time windows for measuring FFs are $50\ ms$. Averages are taken across the excitatory population for the measurements in **(C-F)**. The external input is set as $Q_o = 5\ Hz$ in **(B-F)**.

## 3.3 Scale-free neuronal avalanches near the critical transition point

For our mean-field theory of balanced networks, the Hopf bifurcation predicts that a global oscillation emerges in the field equations with oscillation amplitude growing continuously from zero. This picture is similar to a second-order phase transition in statistical physics when oscillation amplitude is taken as the order parameter. Thus, we measure the spiking avalanches (see Materials and Methods for further details) to examine whether this kind of criticality can result in scale-invariant spiking behaviour, i.e. critical avalanches, as observed in experiments (Beggs and Plenz, 2003; Friedman et al., 2012; Bellay et al., 2015; Fontenele et al., 2019). **Figure 3B** illustrates the time course of an avalanche.

To more clearly compare the feature of avalanche dynamics, raster plots of the spiking time of excitatory neurons in three typical cases are shown in **Figure 3A**. These plots illustrate the asynchronous state (upper panel), the onset of synchrony and oscillation (middle panel) and the developed collective oscillation (lower panel). Power spectrum density analysis of network firing rate is shown in **Figure 3D**. The asynchronous state appears without collective oscillation (frequency peak) and the avalanche size and duration distributions are exponential-like, as shown by the blue curves in **Figure 3E, F**. It is thus subcritical. Alternatively, the asynchronous state can be considered as very noisy population oscillations that are participated by neurons very sparsely. The onset of synchrony is usually accompanied by a frequency peak of fast oscillation situated within the gamma bands, which is thought to have functional importance in various cognitive processes (Herrmann et al., 2004). This fast oscillation is temporally organized as scale-invariant avalanches, with power-law-like size and duration distributions, as shown by the green curves in **Figure 3E, F** and it is thus a critical state. It can be seen from **Figure 2C** that the average pairwise correlation is still low in the critical state, since the avalanches at this stage are only randomly participated by a small portion of neurons (**Figure 2E**), which reconciles the coexistence of weak pairwise correlation and strong clustered spiking patterns (Schneidman et al., 2006). The collective oscillation state has more slow frequency peaks and the avalanche size and duration distributions have heavier tails (corresponding to the red curves in **Figure 3E, F**) compared with the critical state, which are features of a supercritical state. This is because avalanches from collective oscillations with specific harmony can produce typical scales, giving rise to heavy tail in the avalanche size or duration distribution.

We examine the distance $D$ between the avalanche size distribution and its fitted power-law distribution (see Materials and Methods for further details). **Figure 3D** shows that the network was closest to criticality when poised near the Hopf bifurcation point predicted by the mean-field theory. The existence of power-law distribution is only partial evidence of criticality, as other mechanisms could generate power-law distribution (Yadav et al., 2017). We further examined the scaling relation (Sethna et al., 2001) in the critical state (see Materials and Methods for further details). The estimated slopes within the truncated ranges in the avalanche size and duration distributions define the critical exponents $P(S) \sim S^{-\tau}$ and $P(T) \sim T^{-\alpha}$. A third exponent is defined as $\langle S \rangle(T) \sim T^{1/\sigma \nu z}$, where $\langle S \rangle(T)$ is the average size of avalanches with the same duration $T$. The third scaling feature can be found in both the subcritical and critical cases (**Figure 3G**), in accordance with previous experimental findings (Friedman et al., 2012). We find that the scaling relation (Sethna et al., 2001)

$$\frac{\alpha-1}{\tau-1} = \frac{1}{\sigma\nu z} \tag{16}$$

approximately holds at the critical state. The exact value of critical exponents may depend on the details of the system. To demonstrate this, we slightly vary the input strength $Q_o$ from 4 to 8 $Hz$ and the network still shows significant critical properties at the critical value $\tau_d^I = 3\ ms$. As can be seen in **Figure 3H**, different exponents keeping the scaling relation Equation (16) can be found by varying the input strength $Q_o$. This phenomenon, together with the approximate linear relationship between exponents $\tau$ and $\alpha$ (inset of **Figure 3H**), is in accordance with the results measured *in vivo* in the primary visual cortexes (V1) of various animals (Fontenele et al., 2019) across a wide range of neural activity states, as well as our experimental data analysis later. The scaling relation expressed by Equation (16) provides additional evidence that the avalanches in the microscopic network occurring near the bifurcation point of the mean-field model possess the properties of criticality.

Note that an avalanche can be understood as a temporal propagation of spiking activity in a network. These temporal propagations occur when the excitatory current temporally spill over the network and the avalanche size depends on the strength of the propagations, referring to **Figure 1H, I**. In the strict balanced state where excitation is cancelled by inhibition very fast, only small size avalanches can occur and the avalanche distribution is thus subcritical. On the contrary, in the loose balanced state with sparse network oscillations, large avalanches with typical scales can be induced by the temporal domination of excitation and the avalanche distribution can become supercritical. Only near the transition point where the macroscopic dynamic is also noise-driven, avalanches occur with all scales and the avalanche distribution thus can be scale-free. More specifically, on the macroscopic scale, the dynamical process of avalanche corresponds to a noise-induced excursion of the population firing rate. As our measurement of avalanches in the network by individual spiking times requires a fine time scale, and the information in this fine scale is averaged out in the field model, thus, information describing small avalanches vanishes in the field model, which only predicts the global firing rate dynamics. Although it is still difficult to directly link the microscopic avalanches dynamics to the macroscopic firing rate dynamics, the scale-invariance property of criticality inspires that the properties of burst activities (avalanche) in the macroscopic scale can shed light on the origin of power-law scaling in the microscopic network.

To consider the avalanche in the macroscopic scale, one should inspect the fluctuation behavior of a macroscopic signal $x(t)$ of the network, such as the mean firing rate, etc. An avalanche is a process that starts growing at $x(0) = x_{th} + \varepsilon\ (\varepsilon \to 0^+)$ and at the first time it goes back to $x(T) = x_{th}$ at time $t = T$. Here, $x_{th}$ is a threshold above which the avalanche is defined. Such an approach is often used to measure neural avalanche in macroscopic signal such as magnetoencephalography (MEG) (Shriki et al., 2013). The quantity $T$ turns out to be the first-passage time (FPT) (back to $x_{th}$) of this process and it defines the duration of an avalanche. The area $S = \int_0^T x(t)dt$ in this process defines the size of the avalanche. The scale-free behavior at the synchronous transition point may be understood as the general feature of dynamical systems near a bifurcation point when subjected to demographic noise $\frac{dx}{dt} \sim \sqrt{x}\xi(t)$ (di Santo et al., 2017). This is because in the critical state of the E-I network, neurons are subjected to Poisson-like noise input with very weak correlation. Thus, according to a Gaussian approximation given by the central limit theorem, the overall fluctuation of population activity density scales with its square root as given by the central limit theorem (Benayoun et al., 2010; di Santo et al., 2018). Noise-driven random walker theory predicts that power-law distribution of the avalanche is a general feature of a dynamical system subjected to such kind of noise when near the Hopf bifurcation (actually, general bifurcations) point. Specifically, although the dynamic form of this macroscopic signal may not be explicit in the field equations, we can heuristically consider a situation that $X(t) = x - x_{th}$ obeys an intrinsic dynamic as the normal form of the amplitude dynamics of Hopf bifurcation but driven by noisy force modeled as GWN. The Langevin equation it obeys is

$$\frac{dX}{dt} = aX - X^3 + \eta(X, t) \,. \tag{19}$$

The first part of Equation (19) is the normal form of the oscillation amplitude dynamics of a supercritical Hopf bifurcation (Marsden and McCracken, 2012), where periodic motion arises when $a$ increases across the bifurcation point $a = 0$. The second part $\eta(X, t)$ is the noisy driving force, where the fluctuation scales with square root of the activity. Thus, it has the form $\eta(X, t) = h + \sqrt{X}\xi(t)$, where $h$ is the mean bias, including the effect from recurrent excitatory, recurrent inhibitory and external inputs. $\xi(t)$ is a standard GWN. The fact is that avalanche dynamics given by the first-passage process of Equation (19) can be mapped to the case of random walks in logarithmic potential (di Santo et al., 2017) by a scaling analysis (details in (di Santo et al., 2017)). Under this approach, the FPT distribution of the avalanche process can be solved by absorbing boundary approach in an analytical way, resulting in

$$P(T = t) = \frac{(\sqrt{2}\varepsilon)^{1-4h}}{\Gamma(\frac{1}{2}-h)} t^{\frac{4h-3}{2}} \exp(-\frac{2\varepsilon^2}{t}) \,. \tag{20}$$

This explains the power-law distribution relation $P(T) \sim T^{-\alpha}$ with $\alpha = \frac{3}{2} - 2h$ and other exponents can be obtained by scaling argument (di Santo et al., 2017), although the relation between the avalanche critical exponents in the macroscopic scale and the microscopic scale needs further exploration. In all, irregular microscopic spiking leads to macroscopic fluctuation, which becomes the dominant effect that shapes the dynamic when near the macroscopic bifurcation point. The scale-free avalanche dynamics in the microscopic spiking network at the critical state may be understood as the scaling features of the first-passage dynamics near the Hopf bifurcation point in the macroscopic field model.

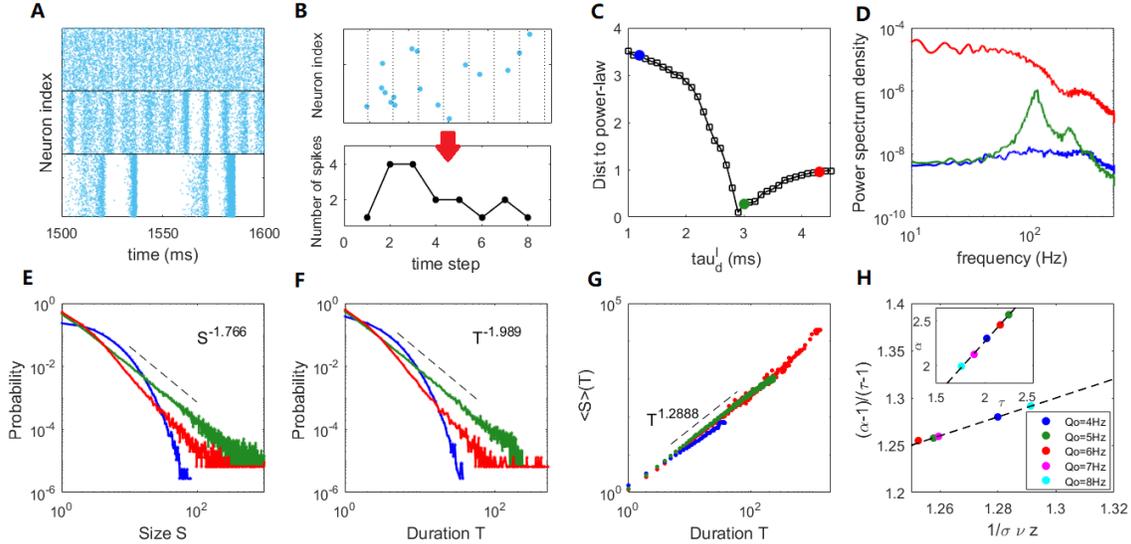

**Figure 3. Critical dynamics near the onset of collective oscillation.** (**A**) The raster plots of the spiking time of the excitatory neurons at $\tau_d^I = 1.2\ ms$ (upper panel), $\tau_d^I = 3\ ms$ (middle panel) and $\tau_d^I = 4.3\ ms$ (lower panel). (**B**) An example of the time course of an avalanche. (**C**) The distance between the avalanche size distribution and its best-fitting power-law distribution. (**D**) The power spectrum density of the network excitatory firing rate. (**E**) The probability density distribution of the avalanche size. (**F**) The probability density distribution of the avalanche duration. (**G**) The mean avalanche size with respect to a given avalanche duration. For (**D-G**), the blue, green and red curves correspond to the cases of $\tau_d^I = 1.2, 3, 4.3\ ms$, respectively. (**H**) Scaling relation Equation (16) approximately holds for critical states ($\tau_d^I = 3\ ms$) with different input strengths $Q_o$, where the dashed line represents Equation (16) exactly holding. The inset shows an approximate linear relationship between exponents $\tau$ and $\alpha$. Here, $Q_o = 5\ Hz$ is used in (**A-D**) and the example of (**E-G**) is $Q_o = 8\ Hz$.

## 3.4 Coexistence of irregular spiking and critical avalanches in experimental data

In the following, we further analyze public experimental data measuring the *in vitro* neuronal spiking activity of mouse somatosensory cortex cultures (Ito et al., 2016) (25 data sets in total) to seek for the phenomena in the E-I balanced model. For the experimental data, spiking in the culture clearly shows up-down state transition. Active spiking periods (up-state) and silent periods (down-state) alternate slowly with a frequency of *circa* 0.1 Hz. We focus on analyzing the up-state (see the definition in Materials and Methods) for the following reasons. First, neurons in up-states exhibit more spiking and stronger oscillation trends, which can be seen from the power spectral density (PSD) in **Figure 4B**. On the contrary, spikes were too few in the down-state, thus up-states are more likely to represent normal neural activities and closer to the dynamic regime we have studied in the model. Furthermore, the neurons recorded in data exhibit heterogeneity (Ito et al., 2014), i.e. board distribution of individual firing rate, which is different from our modeling results with homogeneous random network topology. However, the spiking of neurons in up-state is more homogeneous, which can be seen from the evener distribution of firing rates of neurons (**Figure 4A**). Taken together, we expect that our modeling results may explain partial properties of the experimental data in up-states.

While most of the data sets display heavy tails in the distributions of avalanche size and duration (Supplementary Figure 3 and Supplementary Figure 4), we find that nine of the 25 data sets exhibit both size and duration distributions that correspond to power-laws according to our standards (see Materials and Methods for further details). Among them, seven sets contain critical exponents that approximately satisfy the scaling relation Equation (16) with errors < 0.1. Further details of the analysis results of the data sets are presented in Supplementary Table 1. Estimating the critical exponents of the data sets. These estimated critical exponents $\alpha$, $\tau$ and $\frac{1}{\sigma\nu z}$ are also close to the ranges found in our model. The statistical results of these seven data sets are shown in **Figure 4C, G**. The boxplots of the distribution of CV of ISI across neurons in each set are shown in **Figure 4C**. Typical values of CV of neuron ISI are around 1~1.5, a range similar to the model prediction, indicating irregular spiking. **Figure 4D to F** illustrates the avalanche size and duration statistics of Set 1. Power-law distribution of avalanche size and duration can be observed. For surrogated data, where the inter-spike intervals of neurons were randomly shuffled, power-law distributions cannot be maintained, indicating that the critical properties are intrinsic in the data. As shown in **Figure 4G**, the scaling relationship between critical exponents (i.e. Equation (16)) approximately holds for these data sets. For these critical data sets, we also find a linear relationship between exponents $\tau$ and $\alpha$ (as seen in the inner panel of **Figure 4G**) whereby $\alpha = 1.36\tau - 0.37$, while the slope is slightly different from the value in our model with $\alpha = 1.21\tau - 0.154$ (inner panel, **Figure 3H**) and a previous study (Fontenele et al., 2019) with $\alpha = 1.28\tau - 0.28$. These results, obtained from experimental data, indicate the phenomena of the model that irregular neuron spiking can collectively organize as scale-invariant critical avalanches are widespread.

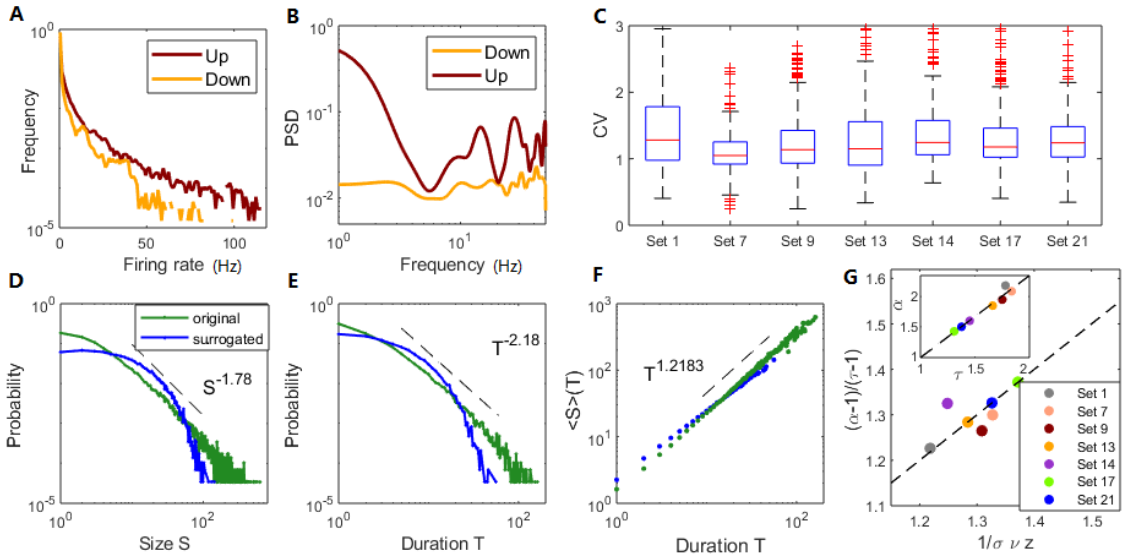

**Figure 4. Coexistence of irregular spiking and critical avalanches in experimental data sets.** **(A)** The frequency distribution of the neurons firing rate in up and down states. **(B)** The power spectral density of the population firing rate in up and down states. In **(A, B)**, the examples are shown by an up and a down period of data Set 1. **(C)** Boxplots of the CV distribution across neurons in different data sets. **(D-F)**: critical avalanche properties of data Set 1. **(D)** Probability density distribution of the avalanche size. **(E)** Probability density distribution of the avalanche duration. **(F)** Mean avalanche size with respect to the given avalanche duration. The power-law distributions of size and duration are destroyed by surrogated data with shuffled ISI (denoted by blue curves). **(G)** Scaling relations between critical exponents of different data sets, similar to **Figure 3H**.

## 4. Discussion

In summary, we have shown that E-I balanced IF networks with synaptic filtering kinetics can reconcile the coexistence of irregular spiking and collective critical avalanches near a synchronous transition point. The mechanism is unveiled by the macroscopic field equations derived by a novel mean-field theory which effectively capture the network dynamics. Finally, we further show that the phenomenon can be widely observed in the spontaneous spiking activities recorded *in vitro* in mouse somatosensory cortex cultures.

### 4.1 The E-I balanced state reconciles with different levels of synchrony

Traditionally, E-I balance (Okun and Lampl, 2009) has been deemed to be the origin of neuron spiking irregularity. Classical mean-field theory (Van Vreeswijk and Sompolinsky, 1996, 1998; Renart et al., 2010) predicts that irregular spiking is an asynchronous state with low spiking time correlation. However, input synchrony (Stevens and Zador, 1998) has also been shown to contribute to the property of irregular spiking. This varied understanding renders a question on how to effectively characterize the dynamic of the E-I balance state.

Here, we study E-I networks that incorporates synaptic kinetics and the critical transition from strict to looser balance, where individual neurons continue to spike irregularly during these different dynamic states. Thus, individual irregular spiking is compatible with asynchronous (Renart et al., 2010) or sparse synchrony (Brunel and Hakim, 2008) state (i.e. synchronous inputs). Our theory shows that the balanced state can be a stable fixed point or a stable limit cycle on the macroscopic scale. The former induced by strict balance corresponds to the asynchronous state in accordance with the traditional theory, while the latter induced by loose balance corresponds to collective network oscillation, with critical dynamics during the transition between them. The critical dynamics where neurons are weakly synchronous also provides an explanation of how weak pairwise correlation can induce abundant collective behavior (Schneidman et al., 2006). Our study thus gives an effective characterization that the E-I balanced state can be a stable state of different characteristics (fixed point or limit cycle) in the macroscopic scale.

## 4.2 Possible origin of crackling noise in neural systems around synchronous transition point

The criticality of neural systems has been long debated (Wilting and Priesemann, 2019a). Previous experimental and theoretical studies (Beggs and Plenz, 2003; Lombardi et al., 2012; Poil et al., 2012; Yang et al., 2012) have suggested that critical avalanches exist in the E-I balanced state, while many simplified models use unrealistic neural dynamics (such as spreading processes on networks) and the definition of E-I balance is usually ambiguous. Our study of IF network with realistic synaptic kinetics considers realistic irregular asynchronous or synchronous spiking in E-I balanced state, and reveals that that critical avalanches exist near the synchronous transition point between these states.

Compared with well-studied simplified physical models, our discovery offers a more biologically plausible explanation of the origin of scale-free dynamics in neural systems, i.e. the dynamics are caused by a potential phase transition indicated by a Hopf bifurcation on the macroscopic scale of E-I balanced neuronal networks. Our theory is thus consistent with the understanding that criticality occurs at the edge of a synchronous transition (di Santo et al., 2018; Fontenele et al., 2019) with intermediate levels of spiking variability (Fontenele et al., 2019), and that critical avalanches can temporally organize as collective oscillations (Gireesh and Plenz, 2008). Significantly, in our model, the synaptic transmission is not instantaneous due to the filtering effect, which renders difficulty in the distinction of the spatially causal relation (Martinello et al., 2017) between successive spiking. The time binning avalanches we measure here are temporally causal (as in experimental measurements) but not necessarily spatially causal in the network. Thus, our estimated critical exponents do not agree with the spatially causal avalanches produced by critical branching processes (Haldeman and Beggs, 2005) (i.e. directed percolation (DP) class), which predicts the classical results $\tau = 1.5, \alpha = 2, \frac{1}{\sigma v z} = 2$. Indeed, critical exponents different from the DP class have been found in previously reported experiments (Palva et al., 2013; Fontenele et al., 2019). Our own analysis of the experimental data (**Figure 4G**) also confirmed the variation of the exponents while maintaining the scaling relationship Equation (16). Furthermore, the linear relationship between exponents $\tau$ and $\alpha$, also found in recent studies (Dalla Porta and Copelli, 2019; Fontenele et al., 2019), may depend on detailed properties of the underlying circuit and its dynamic origin remains to be further explored.

Unlike our study of dynamical criticality here, another branch of study of neural criticality uses concepts in thermodynamics (Mora et al., 2015; Tkačik et al., 2015) (e.g. define criticality when a pre-defined specific heat estimated from neural activities diverges). It remains to be further studied whether such statistical criticality also emerges near the dynamical critical point in our theory.

### 4.3 A macroscopic description of IF neuronal networks with synaptic kinetics

In traditional theory, the firing rate of an IF neuron can be estimated using statistics relating to the first-passage process of membrane potential with diffusion approximation (Burkitt, 2006). Under the assumption of inputs with Gaussian white noise (without auto-correlation), the firing rate statistics can be derived from the first-passage time theory of the Ornstein-Uhlenbeck (OU) process (Ricciardi and Sacerdote, 1979). However, the effect of synaptic filtering will induce memory effects so that the input noise is no longer white, which prevents a complete analytic treatment of the system (for fast decay synapses, the firing rate has been derived asymptotically through a singular perturbation method) (Fourcaud and Brunel, 2002). Self-consistent relation of steady-state firing rate of a network (Shriki et al., 2003; Renart et al., 2004) can be derived by a presumed f-I curve, which expresses the relationship between input current and output rate.

A more challenging but also more useful issue that has attracted much recent attention is to find macroscopic transient dynamic descriptions of the neuronal networks. Schaffer et al. (Schaffer et al., 2013) derived the complex firing rate equations of IF networks through the eigenfunction expansion of the Fokker-Planck equation under diffusion approximation. Deriving rate equations of adaptive nonlinear IF networks has also been studied (Augustin et al., 2017) under some effective approximation of the Fokker-Planck equation. Montbrio et al. (Montbrió et al., 2015) derived the rate equations for quadratic IF networks using the Lorentzian ansatz. This approach has been generalized to including gap junctions (Laing, 2015) and synaptic filtering kinetics (Dumont and Gutkin, 2019). Schwalger et al (Schwalger et al., 2017) developed a method to derive the stochastic rate equations of adaptive IF networks based on mean-field approximation of the renewal equation. In general, analytical theory works for specific conditions and thus is highly specific and with strong complexity. Furthermore, most proposed theories failed to capture the synchronous transitions induced by the synaptic filtering effect studied in our model.

Instead of being theoretically perfect, here we seek for simplicity and effectiveness. The key feature of our framework (see detailed derivations in Materials and Methods) is that the mean-field equations of the macroscopic dynamical variables can be closed by the voltage-dependent mean firing rate (i.e. Equation (9)). While many previous theoretical analyses of E-I network usually require the assumption of low input correlation (VanVreeswijk and Sompolinsky, 1998; Brunel, 2000), the formula here is constructed by counting the number of neuron spiking in a small time window but not required to explicitly consider the correlation of neurons, which allows it to essentially capture the sub and supra threshold microscopic dynamics of a spiking network. Our novel scheme to derive the dynamic equations governing the first-order statistics of the neural network is not completely analytical, since the derivation neglects several factors that shape the network dynamics, including higher order statistics, noise correlation and refractory time. All these neglected factors have been incorporated into the effective parameters, $\sigma_\alpha$, given by Equation (10). Nonetheless, this semi-analytical theory provides a simple yet useful method to analyze dynamic features related to the network firing rate, such as synchrony, criticality and response to dynamic stimuli. Importantly, the theory successfully explains the dynamic origin of the collective oscillation induced by prolonged inhibitory synaptic decay times in the biologically plausible E-I balanced network. The semi-analytical nature of the theory results in the simple form of the macroscopic field equations, which facilitates further generalization.

## 4.4 Conclusion and Outlook

In summary, our theory extends the traditional understanding of E-I balance, the asynchronous state to more realistic dynamics, with the different levels of synchronization that are typically observed in various experiments. It uncovers a possible origin of criticality in neural systems whereby the Hopf transition point in E-I balanced neural circuits can simultaneously reconcile irregular spiking and critical avalanches, a phenomenon that is further observed with *in vitro* experimental data. Our novel semi-analytical mean-field theory offers a simple yet useful and highly generalizable theoretical tool to analyze the dynamics of integrate-and-fire neuronal networks with realistic synaptic kinetics, building a foundation for the integrated study of neural dynamics on different spatial scales. Thus, it is straightforward that the mean-field analysis introduced here can be used in the study of neural dynamics across a diverse range of topics. It can be easily generalized to include other factors by assuming different macroscopic variables in the field model. Possible extensions of the analysis include studying the effects of multiple neuronal populations and synaptic receptor types, cluster or spatially extended network connections, adaptive behaviors such as short-term plasticity, etc. Thus, our work allows further exploration of the mechanism that determines the role of synaptic kinetics in working memory retrieval (Mongillo et al., 2008), self-organized critical phenomena due to plasticity (Levina et al., 2007; Millman et al., 2010) and spatially causal avalanches or waves (Keane and Gong, 2015; Keane et al., 2018; Gu et al., 2019) arising from competition between Hopf and Turing instability (Huang et al., 2019). As a theory that links microscopic neuronal spiking and macroscopic collective activity that are consistent in several aspects with real neural dynamics with regard to E-I balance and neural criticality, our theory also establish a base to model large-scale brain connectomes (Haimovici et al., 2013; Chaudhuri et al., 2015; Wang et al., 2019), to study large-scale brain networks and information processing with realistic multiscale complex dynamics. The potential applications of our theory listed here will receive the further exploration they deserve in due course.

## Funding

This work was supported by the Hong Kong Baptist University (HKBU) Strategic Development Fund, the Hong Kong Research Grant Council (GRF12200217), the HKBU Research Committee and Interdisciplinary Research Clusters Matching Scheme 2018/19 (RC-IRCMs/18-19/SCI01), the National Natural Science Foundation of China (Grants No. 11775314, No. 91530320 and No. 11975194) and Key-Area Research and Development Program of Guangzhou (Grants No. 202007030004). This research was conducted using the resources of the High-Performance Computing Cluster Centre at HKBU, which receives funding from the RGC and the HKBU.

Xue, M., Atallah, B. V, and Scanziani, M. (2014). Equalizing excitation--inhibition ratios across visual cortical neurons. *Nature* 511, 596.

Yadav, A. C., Ramaswamy, R., and Dhar, D. (2017). General mechanism for the 1/f noise. *Phys. Rev. E* 96, 22215.

Yang, D.-P., Zhou, H.-J., and Zhou, C. (2017). Co-emergence of multi-scale cortical activities of irregular firing, oscillations and avalanches achieves cost-efficient information capacity. *PLoS Comput. Biol.* 13, e1005384.

Yang, H., Shew, W. L., Roy, R., and Plenz, D. (2012). Maximal variability of phase synchrony in cortical networks with neuronal avalanches. *J. Neurosci.* 32, 1061–1072.

Zhou, F.-M., and Hablitz, J. J. (1998). AMPA receptor-mediated EPSCs in rat neocortical layer II/III interneurons have rapid kinetics. *Brain Res.* 780, 166–169.
32

# *Supplementary Material*

## Appendix 1. Model Extensions

**1.1 Extension to time-varying inputs.**

The field model Equations (11), (12) can be conveniently used to compute the transient firing rate dynamics of the network in response to the time-varying external input. For inhomogeneous Poisson external inputs with time-dependent firing rate $Q_o(t)$, the constant term $Q_o$ in Equation (12) can just be modified to the time-dependent term $Q_o(t)$ and the field model can be simulated directly in this way. Note that a good estimation of the effective parameter $\sigma_E, \sigma_I$ may depend on $Q_o$, as the estimation result given by Equation (10) depends on $Q_o$. However, we point out that if $Q_o(t)$ does not change very strongly, the parameters $\sigma_E, \sigma_I$ can be kept constant throughout the change of $Q_o(t)$ once they have been estimated. To demonstrate this, we set a time-varying input

$$Q_o(t) = \begin{cases} \lambda, & t \in [0, 200) \\ 4\lambda, & t \in [200, 400) \\ 2\lambda, & t \in [400, 500) \\ 2\lambda(1 + \sin[\frac{\pi}{100}(t - 500)]), & t \in [500, 800] \end{cases} \tag{S1}$$

with $\lambda = 5Hz$ for each neuron as shown in **Supplementary Figure S1A**. Here, we do not consider the synaptic transmission delay and the synaptic decay times are set as $\tau_d^E = \tau_d^I = 4\ ms$, $\tau_r = 0\ ms$, where the network would be in asynchronous dynamics. Note that the external input $Q_o(t)$ contains constant part, discontinuous jumps and continuous changes. Simulations show that the firing rate changes accordingly respondent to the change of input, as can be seen from the raster plot in **Supplementary Figure S1B**. At the same time, we can simulate the field model Equations (11), (12) with the same time-varying input $Q_o(t)$ and with fixed parameters $\sigma_E, \sigma_I$ used in **Figure 2A**. As shown in **Supplementary Figure S1C**, the field model predicts the changing trend of the average firing rate of the network. It should be noted that this simple scheme ignores some complex nature of the firing rate response properties in the presence of synaptic filtering (Moreno-Bote and Parga, 2004; Ledoux and Brunel, 2011).

**1.2 Extension to conductance-based models.**

The present mean-field theory can be directly generalized to conductance-based (COB) model where the postsynaptic inputs received by each neuron depend on the membrane potential of the neuron. Specifically, we further study a COB model with the dynamic equation Equation (1) replaced by

$$\frac{dV_i}{dt} = f_\alpha(V_i) + (V_E^{rev} - V_i)[g_{\alpha o}\sum_{j \in \partial_i^o} F^E * s_j(t - \tau_l^E) + \tag{S2}$$

$$g_{\alpha E}\sum_{j \in \partial_i^E} F^E * s_j(t - \tau_l^E)] + (V_I^{rev} - V_i)g_{\alpha I}\sum_{j \in \partial_i^I} F^I * s_j(t - \tau_l^I)$$

Here, the reversal potential for excitatory and inhibitory synaptic currents are $V_E^{rev} = 0\ mV$ and

$V_I^{rev} = -70\ mV$ respectively. The synaptic strengths of conductance are set as $g_{EO} = 0.025$, $g_{IO} = 0.04$, $g_{EE} = 0.02$, $g_{IE} = 0.04$, $g_{EI} = 0.27$, $g_{II} = 0.48$. Other notations, parameters and settings are the same as the current-based (CUB) case. Similar to the CUB model, the COB model shows emergence of collective oscillation induced by slow inhibition. Such a critical transition can also be predicted by our mean-field theory as a Hopf bifurcation, while the derivation of the field equation is slightly different. In the CUB case, the field equation Equation (12) can be obtained by taking the average $\langle . \rangle_\alpha$ of the original equation Equation (1) under mean-field assumption. In the COB case, we still can take the average $\langle . \rangle_\alpha$ of the Equation (S2) under mean-field assumption, but have to proceed with the decoupling assumption that

$$\langle V_i [g_{\alpha o} \sum_{j \in \partial_i^o} F^E * s_j(t-\tau_l^E) + g_{\alpha E} \sum_{j \in \partial_i^E} F^E * s_j(t-\tau_l^E)] \rangle_\alpha \approx \quad (S3)$$

$$\langle V_i \rangle_\alpha \langle g_{\alpha o} \sum_{j \in \partial_i^o} F^E * s_j(t-\tau_l^E) + g_{\alpha E} \sum_{j \in \partial_i^E} F^E * s_j(t-\tau_l^E) \rangle_\alpha$$

This is based on the fact that in an E-I balanced network where neurons spike irregularly, one expects that at any given time $t$, the correlation between the membrane potential and the recurrent E, I conductance input for different neurons is small. As such, we get the field equations

$$\frac{dV_\alpha}{dt} = f_\alpha(V_\alpha) + (V_E^{rev} - V_\alpha)\left[g_{\alpha o}\left(n_o Q_o + \sqrt{\frac{n_o Q_o}{N_\alpha}} \xi_\alpha(t)\right) + g_{\alpha E}\Phi_E\right] + \quad (S4)$$
$$(V_I^{rev} - V_\alpha)g_{\alpha I}\Phi_I\ ,\alpha = E, I$$

to replace Equation (12), where $\Phi_\alpha(t) = \langle \sum_{j \in \partial_i^\alpha} F^\alpha * s_j(t-\tau_l^\alpha) \rangle_{E,I}$ still obeys Equation (11). Thus, Equations (11), (S4) constitute the field equations of the COB model Equation (S2). The sigmoid relation Equation (9) can still be assumed and $\sigma_E, \sigma_I$ can be estimated in a numerical way through Equation (10). A summarization and comparison between the field equations of CUB model Equation (1) and COB model Equation (S2) is as follows.

CUB:

$$\begin{cases} \frac{dV_\alpha}{dt} = f_\alpha(V_\alpha) + J_{\alpha o}\left(n_o Q_o + \sqrt{\frac{n_o Q_o}{N_\alpha}} \xi_\alpha(t)\right) + J_{\alpha E}\Phi_E + J_{\alpha I}\Phi_I \\ \left(\tau_d^\alpha \frac{d}{dt} + 1\right)\left(\tau_r \frac{d}{dt} + 1\right)\Phi_\alpha = \frac{n_\alpha}{\left[1+\exp\left(\frac{V_{th} - V_\alpha(t-\tau_l^\alpha)}{\sigma_\alpha} \frac{\pi}{\sqrt{3}}\right)\right]}, \alpha = E, I \end{cases} \quad (S5)$$

COB:

$$\begin{cases} \frac{dV_\alpha}{dt} = f_\alpha(V_\alpha) + (V_E^{rev} - V_\alpha)\left[g_{\alpha o}\left(n_o Q_o + \sqrt{\frac{n_o Q_o}{N_\alpha}} \xi_\alpha(t)\right) + g_{\alpha E}\Phi_E\right] + (V_I^{rev} - V_\alpha)g_{\alpha I}\Phi_I \\ \left(\tau_d^\alpha \frac{d}{dt} + 1\right)\left(\tau_r \frac{d}{dt} + 1\right)\Phi_\alpha = \frac{n_\alpha}{\left[1+\exp\left(\frac{V_{th} - V_\alpha(t-\tau_l^\alpha)}{\sigma_\alpha} \frac{\pi}{\sqrt{3}}\right)\right]}, \alpha = E, I \end{cases} \quad (S6)$$

The calculation of the steady-state and its stability analysis at zero transmission delays can be performed in the same way as in CUB model. The qualitative results are similar to the CUB model. There is a critical value $\tau_d^{I*}$ such that when $\tau_d^I < \tau_d^{I*}$ the steady-state is a stable focus, corresponding to the asynchronous strict balance state of the network. For $\tau_d^I > \tau_d^{I*}$, the steady-state destabilizes through a supercritical Hopf bifurcation (**Supplementary Figure S1D**), corresponding to the onset of collective oscillation in the network, as shown by the PCC in **Supplementary Figure S1E**. The spiking of individual neurons are still irregular, as can be seen from the high CV of ISIs in **Supplementary Figure S1F**. Furthermore, near the Hopf bifurcation point, the COB model exhibits critical properties in terms of avalanche dynamics similar to the results of CUB model. Overall, the quality of theoretical prediction in the COB case is worse than the CUB case. Indeed, the COB input would lead membrane potential more bias to a Gaussian distribution (Richardson and Gerstner, 2005), an assumption in our derivation. A complete analytical treatment of COB model is very challenging (Renart et al., 2004). However, the semi-analytical mean-field approach here constitutes an effective description of the macroscopic dynamics of E-I network, which has an advantage that it works for both CUB and COB dynamics.

## Supplementary Reference

# Appendix 2. Sensitivity of the Critical Points on the Effective Parameters

The mean-field scheme to derive the field equations introduces two effective parameters $\sigma_E, \sigma_I$ to construct the voltage-dependent firing rate relation Equation (9) and they are the crucial quantities that determine the quality of the scheme. Thus, it is important to know how the theoretically predicted critical point $\tau_d^{I*}$ depends on the choice of $\sigma_E, \sigma_I$.

Although the critical point in the field model can be thought as a Hopf bifurcation point, the concept of critical point is not decisive in the E-I spiking neuronal network. This can be seen from **Figure 2C** and **Supplementary Figure S1B** which show that the spiking correlation increases in a somewhat continuous way as $\tau_d^I$ increases. Furthermore, the critical properties of avalanche shown in **Figure 3** are statistical properties so that one can expect that for parameters close to the critical value these properties can still maintain in a statistically significant manner. As a first approximation, the critical point in the E-I spiking network can be defined as the parameter value where the distance of the avalanche size distribution to its best fit power law distribution, $D$ defined in Methods, is minimal, as shown in **Figure 3C**. We find that for large network size, if $\sigma_E, \sigma_I$ are estimated in the numerical way through Equation (10), the critical point in the spiking network, is very close to the Hopf bifurcation point in the field model. We denote $\tau_d^{I*}$ as the Hopf bifurcation point in this 'optimal' estimation of the parameters $\sigma_E, \sigma_I$ using Equation (10).

We compute the Hopf bifurcation point $\tau_d^{I\ Hopf}(\sigma_E, \sigma_I)$ predicted by the field model for different values of $\sigma_E, \sigma_I$ and compare it to the 'real' critical point (estimated by $\tau_d^{I*}$). The difference $\tau_d^{I\ Hopf}(\sigma_E, \sigma_I) - \tau_d^{I*}$ of the CUB model and the COB model can be seen in **Supplementary Figure S2**. From **Supplementary Figure S2**, we can see that once $\sigma_E, \sigma_I$ are estimated with suitable values, such as by Equation (10), the prediction of the critical synchronous transition point is very precise.

We also notice that the bifurcation point predict by the field model seems to mainly depend on the difference $\sigma_E - \sigma_I$. Once $\sigma_E - \sigma_I$ lies on suitable ranges, the predicted critical point will be very close to the 'real' one. It can also be noticed that in the CUB model the critical point is not sensitive to the values of $\sigma_E, \sigma_I$ compared with the COB model, as can be seen from **Supplementary Figure S2A to C** that the difference $\tau_d^{I\ Hopf}(\sigma_E, \sigma_I) - \tau_d^{I*}$ is still low for a large range of parameter values. On the contrary, the sensitivity in the COB case implies that the COB model has more complicated intrinsic dynamic nature that has to be further explored.

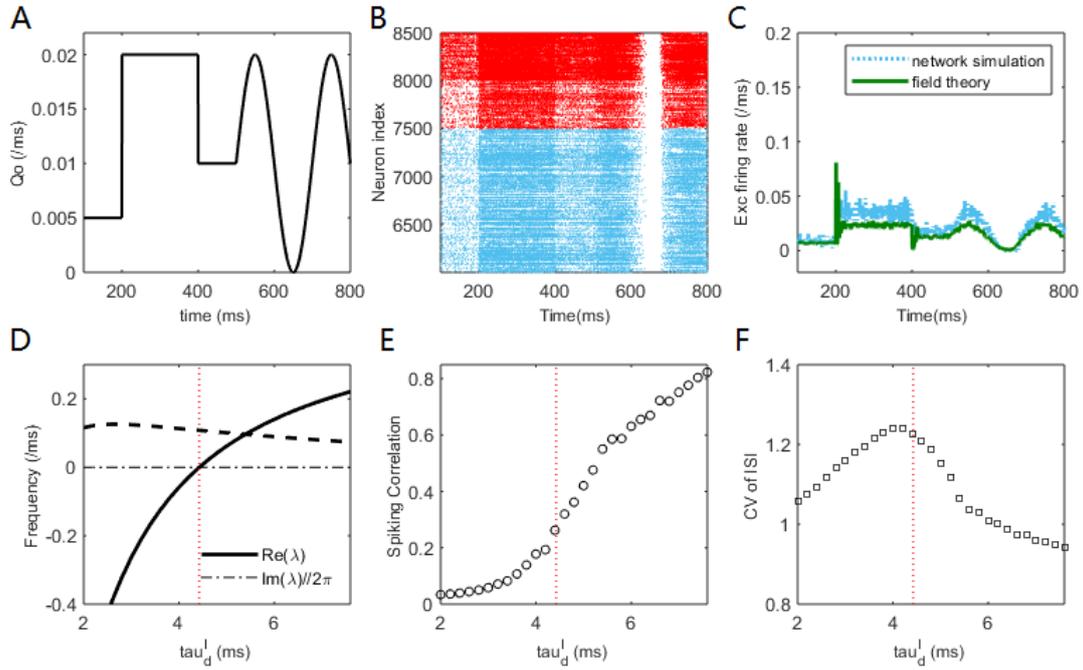

**Supplementary Figure 1. Results of the generalized models. (A-C)** Network dynamics in response to time-varying external input. **(A)** The time-dependent input function $Q_o(t)$ used in simulation. **(B)** Raster plot of the spiking time of neurons (only part of the $N=10000$ neurons are shown). The excitatory/ inhibitory neurons are indicated in blue/red. **(C)** Comparison of the mean firing rate of excitatory population obtained by network simulation and field model simulation. **(D-F)** Mean-field theory prediction of the transition from asynchronous to synchronous state in COB model. **(D)** Field equations predict that a Hopf bifurcation occurs as the increase of inhibitory decay time $\tau_d^I$ at a critical value around $\tau_d^I \approx 4.4\ ms$. The real and imaginary part (divided by $2\pi$) of the dominant eigenvalue are given by the solid and dashed lines, respectively. **(E)** The PCC index shows the emergence of network oscillation as the increase of $\tau_d^I$ across the bifurcation point. **(F)** The CV of ISI at different value of $\tau_d^I$. The parameters in COB model are set as $\tau_l^E = \tau_l^I = 0\ ms$, $\tau_d^E = 2\ ms$, $\tau_r = 0\ ms$ and $Q_o = 5\ Hz$.

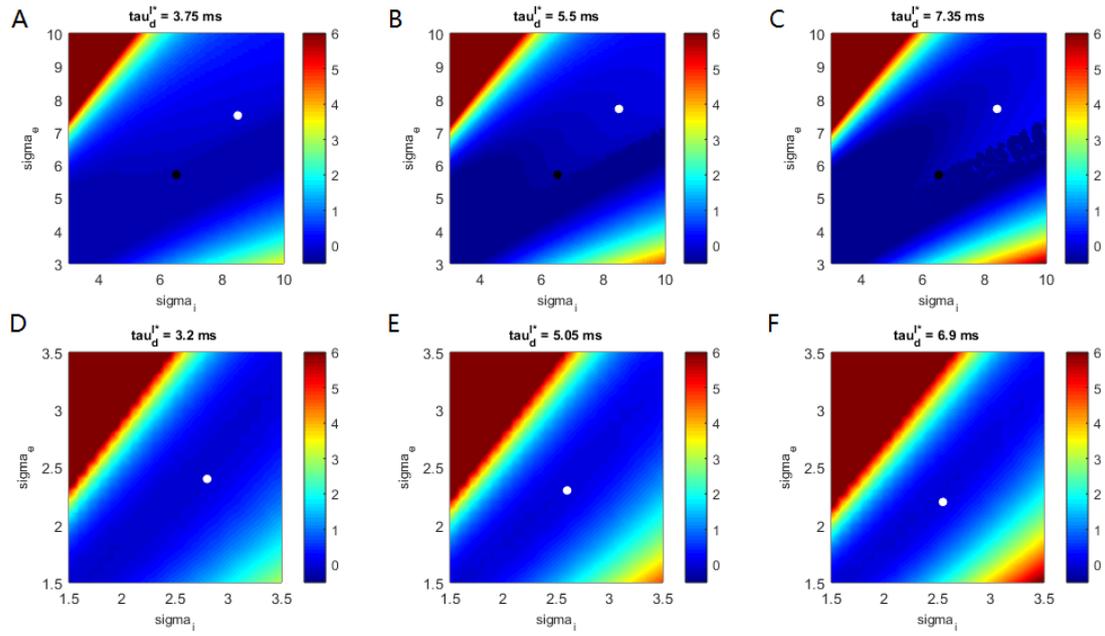

**Supplementary Figure 2. Sensitivity of the predicted critical point on the effective parameters.** The difference value $\tau_d^{I\ Hopf}(\sigma_E, \sigma_I) - \tau_d^{I*}$ is shown by color for current-based (CUB) model **(A-C)** and conductance-based (COB) model **(D-F)**. White dots in **(A-F)** indicate the positions of the numerical estimation results by Equation (10) and the corresponding estimated critical value $\tau_d^{I*}$ is shown on top of the plots. Black dots in **(A-C)** indicate the positions of the theoretical estimation results by fixed $\sigma_E, \sigma_I$ used in **Figure 2A** in the CUB model. Parameters are $\tau_d^E = 2\ ms$ for **(A, D)**, $\tau_d^E = 3\ ms$ for **(B, E)**, $\tau_d^E = 4\ ms$ for **(C, F)** and $Q_o = 5\ Hz$ for all the cases.

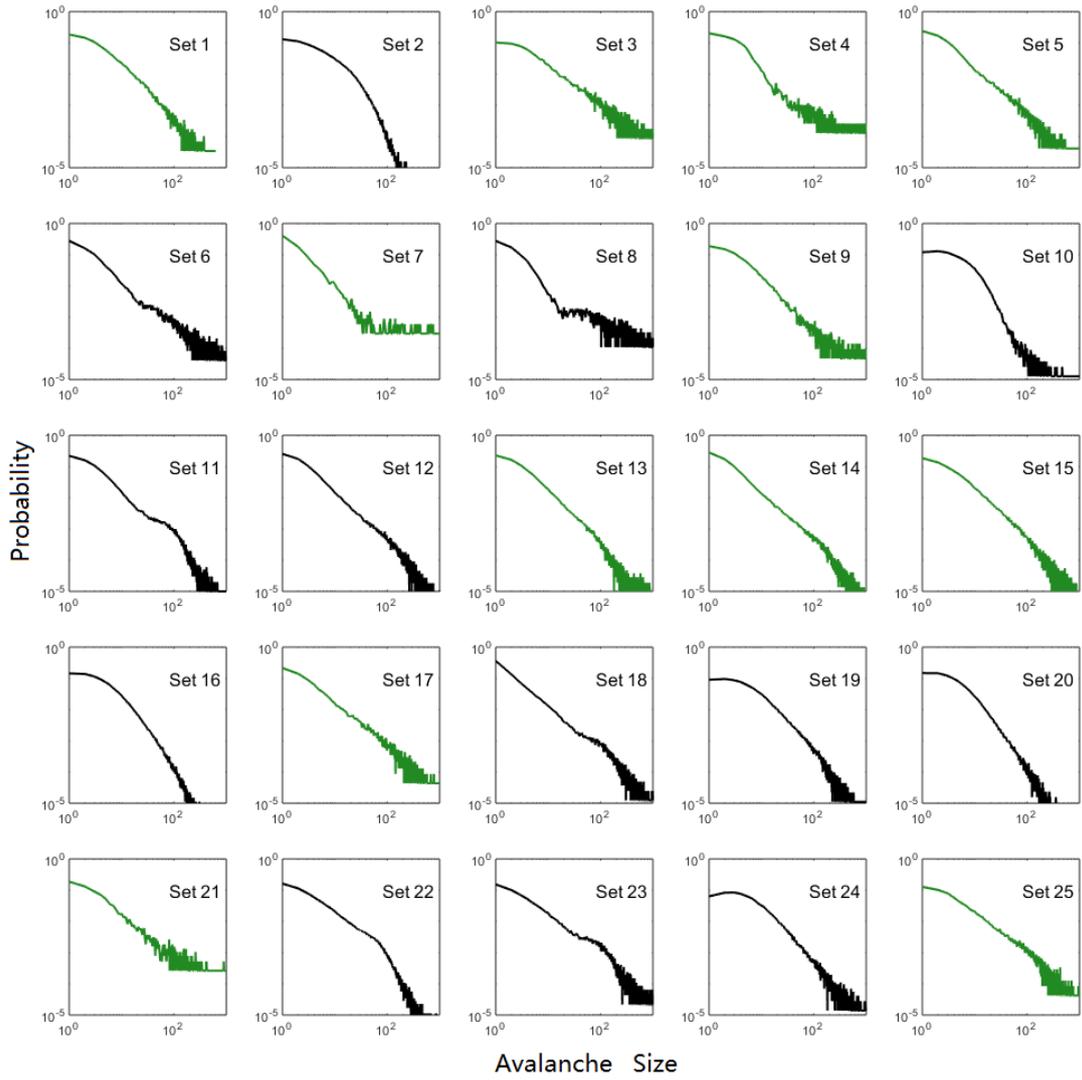

**Supplementary Figure 3. Avalanche size distributions of the data sets.** According to our standard, the up-state of data Set 1,3,4,5,7,9,13,14,15,17,21,25, plotted by green color, exhibit significant power-law size distribution $P(S) \sim S^{-\tau}$. Refer to **Supplementary Table 1** for Details.

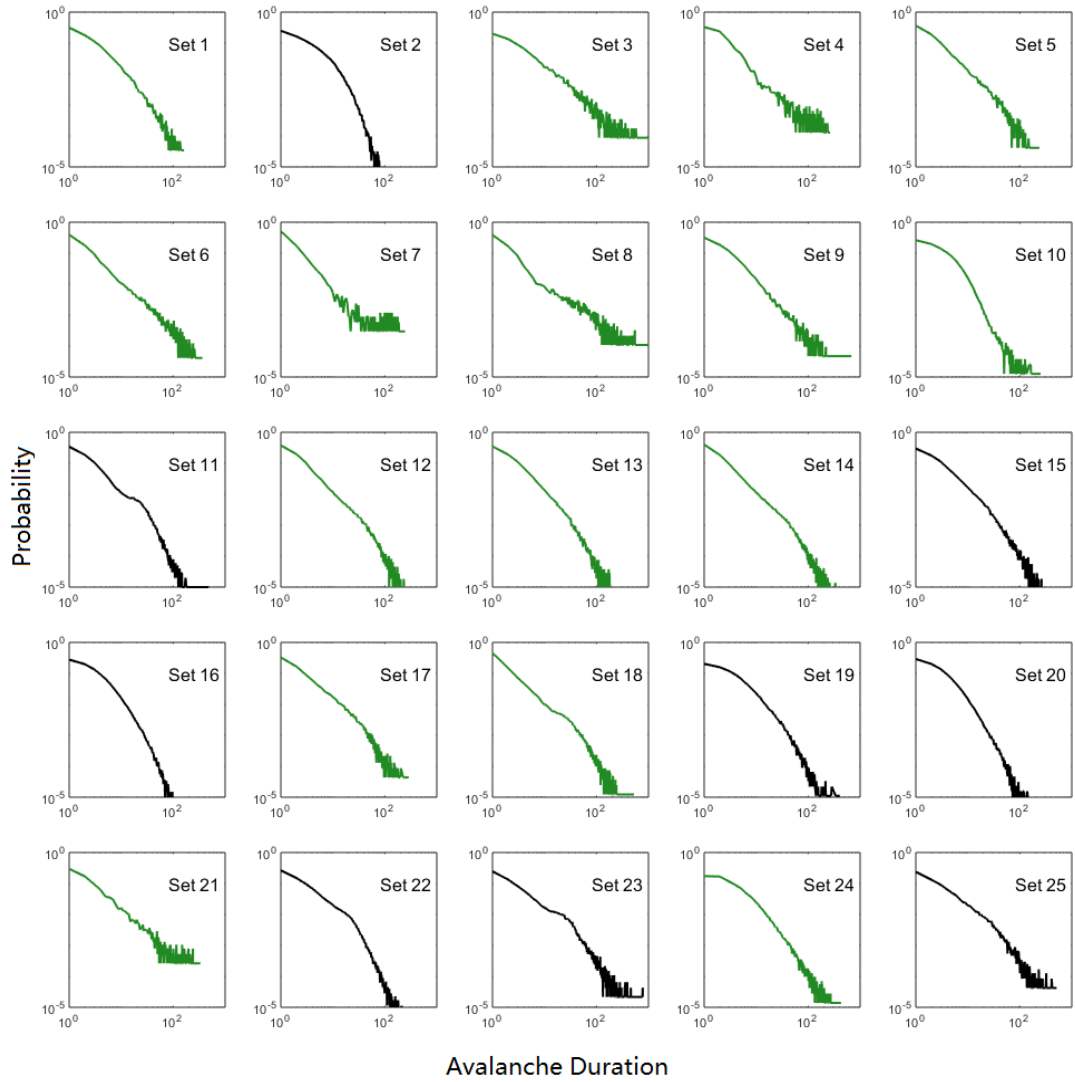

**Supplementary Figure 4. Avalanche duration distributions of the data sets.** According to our standard, the up-state of data Set 1,3,4,5,6,7,8,9,10,12,13,14,17,18,21,24, plotted by green color, exhibit significant power-law duration distribution $P(T)\sim T^{-\alpha}$. Refer to **Supplementary Table 1** for Details.

**Supplementary Table 1. Estimating the critical exponents of the data sets.** The number of neurons, length of the up-states, the time bin used in measuring the avalanches, maximum avalanche size and duration, those estimated critical exponents, data ranges after truncations and the p-values in KS test of the fitted power laws in each data set are shown. Here, the avalanche size and duration are in their original linear scale.

| Set No. | Neuron amounts | Up-state length (mins) | bin size (ms) | max size | max duration | size range | τ | p value | duration range | α | p value | 1/σνz | (α-1)/(τ-1) |
|---|---|---|---|---|---|---|---|---|---|---|---|---|---|
| 1 | 166 | 12.60 | 3.70 | 627 | 162 | [10,95] | 1.78 | 0.118 | [9,50] | 2.183 | 0.108 | 1.218 | 1.226 |
| 2 | 443 | 28.24 | 2.05 | 476 | 115 | | | | | | | | |
| 3 | 99 | 9.16 | 1.97 | 11502 | 1970 | [4,125] | 1.248 | 0.42 | [4,52] | 1.365 | 0.406 | 1.245 | 1.472 |
| 4 | 98 | 8.08 | 5.62 | 1713 | 264 | [12,144] | 1.223 | 0.18 | [11,71] | 1.349 | 0.516 | 1.223 | 1.565 |
| 5 | 367 | 11.35 | 3.53 | 1090 | 237 | [7,124] | 1.419 | 0.188 | [4,25] | 1.644 | 0.846 | 1.321 | 1.537 |
| 6 | 286 | 10.06 | 2.20 | 1798 | 365 | | | | [7,52] | 1.442 | 0.88 | 1.271 | |
| 7 | 172 | 4.90 | 6.43 | 2230 | 245 | [2,27] | 1.837 | 0.984 | [2,14] | 2.088 | 0.804 | 1.327 | 1.300 |
| 8 | 191 | 8.06 | 2.19 | 8002 | 1304 | | | | [6,116] | 1 | 0.644 | 1.204 | |
| 9 | 333 | 7.51 | 2.37 | 2775 | 677 | [5,105] | 1.751 | 0.212 | [5,44] | 1.95 | 0.878 | 1.309 | 1.265 |
| 10 | 368 | 17.56 | 2.44 | 1222 | 251 | | | | [9,59] | 3.431 | 0.114 | 1.366 | |
| 11 | 243 | 16.79 | 1.25 | 2195 | 481 | | | | | | | | |
| 12 | 346 | 14.45 | 1.03 | 1534 | 349 | | | | [7,50] | 1.653 | 0.252 | 1.266 | |
| 13 | 381 | 16.40 | 1.34 | 1409 | 321 | [7,116] | 1.662 | 0.334 | [4,33] | 1.85 | 0.164 | 1.284 | 1.284 |
| 14 | 304 | 16.24 | 0.79 | 2071 | 450 | [7,91] | 1.452 | 0.226 | [6,46] | 1.599 | 0.22 | 1.248 | 1.325 |
| 15 | 206 | 16.49 | 1.07 | 2066 | 445 | [7,90] | 1.566 | 0.704 | | | | | |
| 16 | 594 | 22.72 | 0.87 | 1230 | 278 | | | | | | | | |
| 17 | 435 | 10.07 | 2.75 | 1566 | 287 | [7,113] | 1.309 | 0.356 | [3,24] | 1.424 | 0.794 | 1.372 | 1.372 |
| 18 | 180 | 14.46 | 1.08 | 3280 | 525 | | | | [2,17] | 1.621 | 0.146 | 1.310 | |
| 19 | 262 | 11.58 | 0.88 | 1685 | 407 | | | | | | | | |
| 20 | 534 | 18.02 | 1.49 | 1318 | 273 | | | | | | | | |
| 21 | 444 | 2.91 | 3.02 | 2985 | 334 | [8,117] | 1.377 | 0.806 | [2,53] | 1.5 | 0.978 | 1.326 | 1.326 |
| 22 | 384 | 20.85 | 0.76 | 2598 | 487 | | | | | | | | |
| 23 | 310 | 13.79 | 1.14 | 5015 | 779 | | | | | | | | |
| 24 | 358 | 12.33 | 1.08 | 2093 | 429 | | | | [13,100] | 2.306 | 0.168 | 1.235 | |
| 25 | 231 | 10.71 | 1.93 | 3256 | 510 | [5,150] | 1.229 | 0.632 | | | | | |